%% file: main.tex
\documentclass[%
unsortedaddress,
preprint,
 amsmath,amssymb,
 aps,
]{revtex4-2}

\usepackage{graphicx}
\usepackage{dcolumn}
\usepackage{bm}
\usepackage{subcaption}
\usepackage{comment}
\usepackage{xcolor}
\usepackage{natbib}
\begin{document}

\preprint{APS/123-QED}

\title{Bifurcations and nonlinear dynamics of the follower force model for active filaments}

\author{Bethany Clarke}
\email{b.clarke21@imperial.ac.uk}
\affiliation{%
Department of Mathematics, Imperial College London, SW7 2AZ, UK
}%
 \author{Yongyun Hwang}
 \email{y.hwang@imperial.ac.uk}
 \affiliation{Department of Aeronautics, Imperial College London, SW7 2AZ, UK}
\author{Eric E Keaveny}%
 \email{e.keaveny@imperial.ac.uk}
\affiliation{%
Department of Mathematics, Imperial College London, SW7 2AZ, UK
}%

\date{\today}

\begin{abstract}
Biofilament-motor protein complexes are ubiquitous in biology and drive the transport of cargo vital for many fundamental life processes at the cellular level.  As they move, motor proteins exert compressive forces on the filaments to which they are attached. If the filament is clamped or tethered in some way, this force leads to buckling and a subsequent range of dynamics.  The follower force model, in which a single compressive force is imposed at the filament tip, is a simple filament model that is becoming widely used to describe an elastic filament, such as a microtubule, compressed by a motor protein. Depending on the force value, one can observe different states including whirling, beating and writhing, though the bifurcations giving rise to these states are not completely understood. In this paper, we utilise techniques from computational dynamical systems to determine and characterise these bifurcations. We track emerging time-periodic branches and identify new, quasiperiodic states. We investigate the effect of filament slenderness on the bifurcations and, in doing so, present a comprehensive overview of the dynamics which emerge in the follower force model.
\end{abstract}

\maketitle

\input{intro}

\input{part2.tex}

\input{part3.tex}

\input{part4.tex}
\input{part5.tex}

\input{part6}

\input{part7.tex}

\input{discussion}

\input{appendix}

\bibliography{references}

\end{document}

%% file: intro.tex
\section{Introduction}

Biopolymer filament-motor protein complexes, such as actin-myosin, microtubule (MT)-dynein and MT-kinesin, are essential for driving transport at the cellular level \cite{Cooper2007TheEdition,Needleman2019TheCell}. These systems have importance in cell division \cite{Wojcik2001KinetochoreProtein,Goshima2003TheLine,Forth2017TheDivision}, intracellular transport such as axoplasmic transport in neurons \cite{Holzbaur2011MicrotubulesNeuropathy,Guedes-Dias2019AxonalFunction,Hirokawa2010MolecularDisease} and can, in the case of MT-dynein, collectively form complex superstructures such as the axoneme in cilia
\cite{Gibbons1981CiliaEukaryotes,Gilpin2020TheFlagella}. Cilia control fluid flow at the microscale, for instance to allow for cellular propulsion \cite{Bennett2015AChlamydomonas,Wan2017RunOctoflagellate,Pedley2016SquirmersSwimming, Suarez2006SpermTract}, or can coordinate to pump fluids such as cerebrospinal fluid around the brain \cite{Worthington1966CILIARYSURFACES,Faubel2016Cilia-basedVentricles,Guldbrandsen2014In-depthCSF-PR.,Zhang2015AProteome,Sawamoto2006NewBrain} and mucus in our airways \cite{Hwang1969RheologicalMucus,Lai2009Micro-Mucus}. 

Molecular motors are activated by ATP, which is hydrolised to provide the motor proteins with sufficient energy to carry cargo along a filament \cite{Schliwa2003MolecularMotors,Cho2012TheDomain}. As these motor proteins carry cargo, they experience a drag force from the surrounding fluid. By Newton's third law, they exert a compressive force, equal in magnitude, back on the filament. Experimentally, this force has been found to be in the range 1-10pN \cite{Schliwa2003MolecularMotors,Svoboda1994ForceMolecules,Mallik2004CytoplasmicLoad}. In an unbounded domain, this compressive force would cause the filament to slide in the direction of the force. However if the motion of the filament is restricted in some way, for instance if the filament is attached to a cellular body, or if it reaches an obstacle, this forcing can lead to filament buckling. Such buckling has been observed in motility assays where filaments (actin or MTs) are driven by ATP-fueld motor proteins (myosin or kinesin/dynein respectively) fixed to rigid surfaces \cite{Bourdieu1995SpiralForce,Kawamura2010MicrotubuleMicrotubules,Allen1985GlidingTransport,Gittes1996DirectionalMicrotubule,Young2010DynamicsFlow,Isele-Holder2015Self-propelledDynamics,Lam2014ControllingDensity}.  Undisturbed, the filaments are observed to glide along the surface. However in the presence of an obstacle or surface defect, such as a defective motor protein, part of the filament will become pinned or clamped into place. Then, continued forcing from the motors will cause the filament to buckle. 

In another context, it has been observed that collections of motor-driven MTs can coordinate their motions to generate fluid flows in a process called cytoplasmic streaming \cite{Goldstein2015AStreaming,Niwayama2011HydrodynamicEmbryo,Klughammer2018CytoplasmicContractions}. An example of this occurs within \textit{Drosophila} oocytes \cite{Gutzeit1982Time-lapseDrosophila,Quinlan2016CytoplasmicOocyte,Stein2021SwirlingCytoskeleton}, which are lined internally with kinesin-driven MTs. Kinesin motors induce compressive forces along each MT, which in turn causes each MT to buckle. The dynamics of neighbouring MTs are coupled through the flow within the cell, which allows the filaments to coordinate their motion to generate an overall steady state, in which each MT is bent in the same direction. This directs the fluid around the cell, hence allowing for the transportation of important organelles and molecules.

The mathematical models of motor protein-filament complexes usually treat the MT as an elastic filament driven by internal active stresses that represent the forces generated by the motor proteins. A simple incarnation of such a model considers a single `follower force' at the free-end of a filament that is clamped at its base to a rigid surface \cite{Herrmann1964OnForces,Bayly2016SteadyFlagella,DeCanio2017,Ling2018Instability-drivenMicrofilaments}. The follower force is a compressive force directed along the tangent to the filament centerline and thus changes direction if the filament bends \cite{DeCanio2017,Ling2018Instability-drivenMicrofilaments,Westwood2021CoordinatedSurfaces}. This model neglects the opposite force the motor protein exerts on the surrounding fluid. For an initially straight filament, when the magnitude of the follower force exceeds a critical value, the filament buckles and gives rise to a time dependent state. If motion is restricted to a plane, planar beating \cite{DeCanio2017} is observed, whereas if the filament is allowed to move in three dimensions (3D), buckling gives rise to whirling \cite{Ling2018Instability-drivenMicrofilaments}.  At higher values of the follower force, whirling transitions to planar beating, while for the highest force values, writhing is observed.  This basic follower force model has been extended to distributions of follower forces along the filament length, corresponding to a continuous array of motor proteins \cite{DeCanio2019,Bayly2016SteadyFlagella}, as well as the inclusion of motor motion through an opposite force on the fluid \cite{DeCanio2019,Stein2021SwirlingCytoskeleton}, or a filament surface velocity \cite{Laskar2017FilamentNumber}. Further extensions of the model vary the boundary conditions and initial configuration of the filament \cite{Fily2020BucklingFilaments,Sekimoto1995SymmetrySystem,Fatehiboroujeni2018NonlinearDrag,Fatehiboroujeni2021Three-dimensionalFlipping}, replicating buckling dynamics observed experimentally in motility assays \cite{Fily2020BucklingFilaments,Sekimoto1995SymmetrySystem}. Collections of MTs have been modelled using the follower force model, capturing the collective bending of MTs in cytoplasmic streaming in \textit{Drosophila} oocytes \cite{Stein2021SwirlingCytoskeleton}, and providing a potential mechanism for the onset of ciliary beating \cite{Bayly2016SteadyFlagella,Woodhams2022GenerationModels}.

Despite being used to model complex phenomena, a complete understanding of the basic case of a single filament, clamped at its base to a no-slip planar surface and subject to a follower force at its tip, is still outstanding.  While the different states corresponding to filament whirling, beating, and writhing have been identified, studies have relied primarily on solving initial value problems for different follower force values to determine when these states might arise.  As a result, the precise nature of the bifurcations producing these states and a clear understanding of state stability remain outstanding.  While stability of the trivial, straight filament has been studied quantitatively by assessing the eigenvalues of the linear operator at the bifurcation \cite{DeCanio2017,Ling2018Instability-drivenMicrofilaments}, the differences in emergent states given by the 2D and 3D analyses have yet to be reconciled.  The transition between whirling and beating has not yet been explored, and a characterisation of the complex writhing behaviour observed at higher forcing has not been performed.  Furthermore, previous studies have focused on how the whirling, beating, or writhing vary with the follower force, leaving the dependence of the emergent state on the filament aspect ratio, a key parameter related to the balance of the viscous and elastic forces, largely unexplored.

In this paper, we use techniques from computational dynamical systems to examine in detail the bifurcations that give rise to the beating, whirling and writhing states, exploring how the solutions change as both the strength of the follower force and filament aspect ratio are varied.  In Section \ref{sec:ActiveFilamentModel}, we present the follower force model and describe the numerical methods we use to compute filament dynamics.  We then perform a series of simulations to ascertain the final state achieved for different follower force strengths and filament aspect ratios by allowing the filament to evolve from a straight configuration with a small perturbation.  The resulting state space is provided in Section \ref{sec:ivp}.  In Section \ref{Section:bifurcation_theory} we describe the dynamical systems approach we use to move beyond solving initial value problems and determine directly the time-periodic branches corresponding to whirling and beating.  We ascertain state stability along these branches and categorise the bifurcations of the steady, trivial state and the time-periodic solutions.  Using this approach, we show that buckling of the trivial state is a double Hopf bifurcation and link planar beating found to emerge in 2D to whirling that emerges in fully 3D dynamics.  This is presented in Section \ref{sec:doublehopf}.  In Section \ref{sec:whirlingtobeating}, we explore the transition between whirling and beating and identify a new, quasiperiodic state that connects these two solution branches.  Finally, in Section \ref{sec:beatingtowrithing}, we study the writhing regime and demonstrate that it can be decomposed further into a transitional regime, where a menagerie of quasiperiodic, chaotic and time-periodic solutions are observed, and a regime where only a quasiperiodic solution arises where filament motion can be characterised by a whirling base and a beating tip.  In performing this study, we present a thorough overview of the state space in this fundamental model.

%% file: part2.tex

\section{Follower Force Model}\label{sec:ActiveFilamentModel}

\subsection{Model Description}
\begin{figure}[t]
     \centering
     \includegraphics[width=0.8\textwidth]{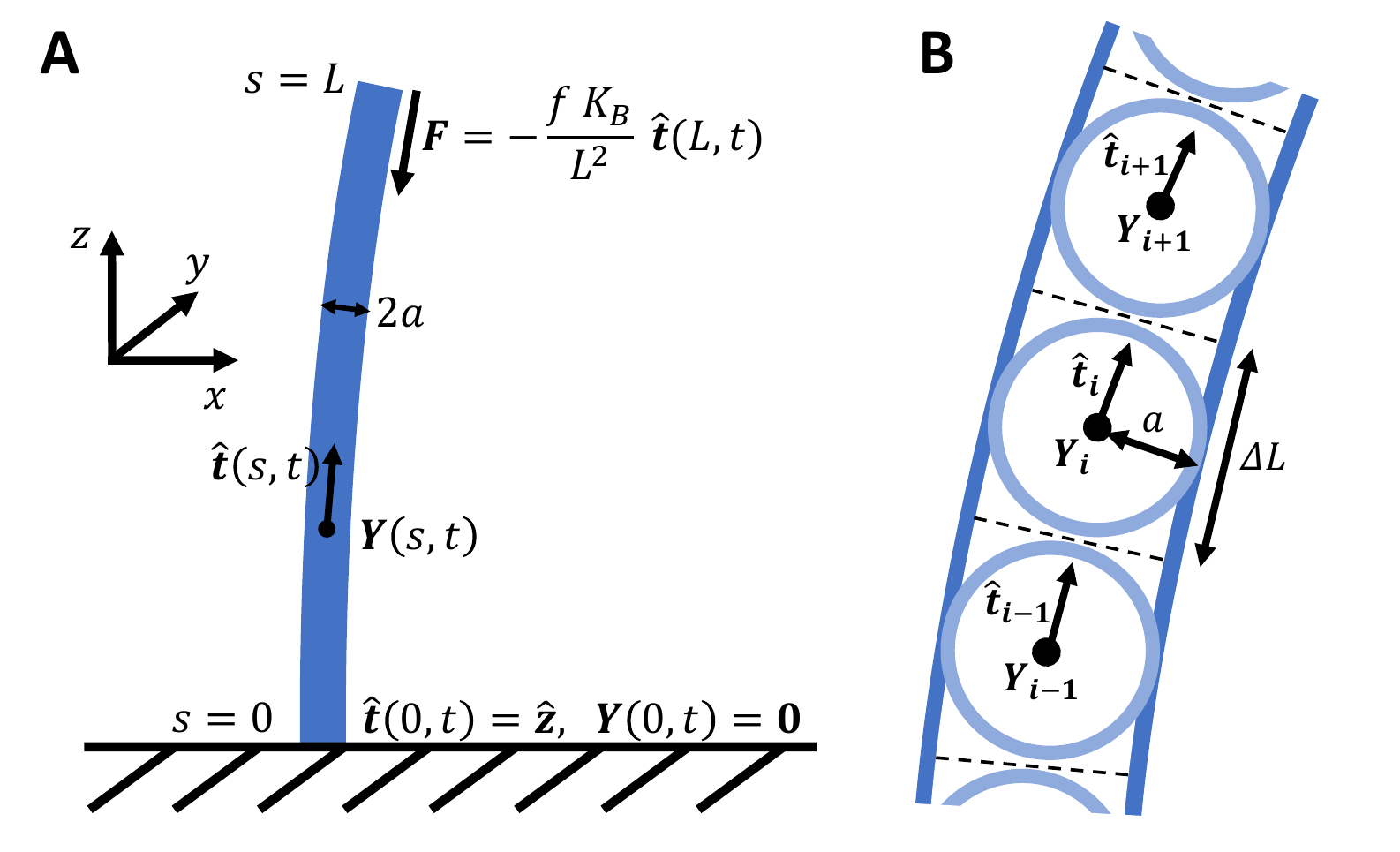}
     \caption{A: Diagram depicting the follower force model. The filament has radius $a$ and length $L$. The bending and twisting moduli are $K_B$ and $K_T$, respectively.  The filament base is clamped to a rigid planar wall at $z=0$ and surrounded by a fluid with viscosity $\eta$. Also shown is the compressive follower force of strength $f$ at the filament's distal end. B: Diagram describing the numerical method used to compute filament evolution.  The filament is discretised into $N$ segments of length $\Delta L$. Segment $i$ has position $\bm{Y}_i$ and local basis $\{\bm{\hat{t}}_i,\bm{\hat{\mu}}_i,\bm{\hat{\nu}}_i\}$ where $\bm{\hat{t}}_i$ is constrained to be the tangent of segment $i$. The hydrodynamic mobility of the segments are given by the RPY tensor with hydrodynamic radius $a$.}
    \label{fig:FF_schematic}
\end{figure}

We begin by describing the model for a filament clamped to a rigid, no-slip planar surface that is driven by a follower force at its free end, as depicted in Figure \ref{fig:FF_schematic}A.  We consider a single filament of length $L$ and cross-sectional radius $a$, such that the filament aspect ratio is given by $\alpha = L/a$. The filament has bending and twisting moduli $K_B$ and $K_T$ respectively, and is surrounded by a fluid of viscosity $\eta$.  As a result of the small velocities and length-scales associated with filament-motor protein systems, the Reynolds number is small and the dynamics can be considered in the overdamped limit.  Accordingly, in the follower force model the effects of fluid and filament inertia are ignored. 

To model the filament dynamics, we employ the methods outlined in \cite{Schoeller2021MethodsFilaments} which we now summarise. The filament centerline position is denoted by $\bm{Y}(s,t)$ which is a function of arc-length, $s\in [0,L]$, and time, $t\in [0,\infty)$ (see Figure \ref{fig:FF_schematic}A). Residing at each point along the filament centerline is the orthonormal local basis $\{\bm{\hat{t}}(s,t),\bm{\hat{\mu}}(s,t),\bm{\hat{\nu}}(s,t)\}$, where $\bm{\hat{t}}(s,t)$ is constrained to be the unit tangent to the filament centerline through
\begin{equation}
     \frac{\partial \bm{Y}}{\partial s} =\bm{\hat{t}}. \label{eq:continuous_equations3}    
\end{equation}

The force and moment balances along the filament are given by
\begin{align}
     \frac{\partial \bm{\Lambda}}{\partial s}+\bm{f}^H &=0, \label{eq:continuous_equations1}\\
     \frac{\partial \bm{M}}{\partial s}+\bm{\bm{\hat{t}}} \times \bm{\Lambda}+\bm{\tau}^H &=0, \label{eq:continuous_equations2}
\end{align}
where $\bm{\Lambda}(s,t)$ and $\bm{M}(s,t)$ are the internal forces and moments on the filament cross-section, and $\bm{f}^H(s,t)$ and $\bm{\tau}^H(s,t)$ are the hydrodynamic forces and torques per unit length. The internal forces enforce the kinematic constraint, (\ref{eq:continuous_equations3}), and the internal moments are given by the constitutive relation \cite{Schoeller2021MethodsFilaments},
\begin{equation}\label{eq:model_moment}
    \bm{M}(s,t) = K_B \left( \hat{\bm{t}} \times \frac{\partial \hat{\bm{t}}}{\partial s}\right) + K_T \left( \bm{\hat{\nu}} \cdot \frac{\partial \bm{\hat{\mu}}}{\partial s} \right) \hat{\bm{t}}.
\end{equation}

The filament is clamped at one end to a infinite, no-slip planar surface located at $z = 0$, fixing the position and tangent of the base as $\bm{Y}(0,t) = \bm{0}$ and $\hat{\bm{t}}(0,t) = \hat{\bm{z}}$, along with the twist angle at the base. At the free end, the moment is zero, $\bm{M}(L,t)=\bm{0}$, and the inclusion of the follower force to the tip provides the condition,
\begin{equation}\label{eq:followerforce_BC}
    \bm{\Lambda}(L,t) = -\frac{f K_B}{L^2} \hat{\bm{t}}(L,t),
\end{equation}
where $f$ is the non-dimensional parameter controlling the magnitude of the follower force.

\subsection{Numerical Discretisation}
Following \cite{Schoeller2021MethodsFilaments}, we discretise the filament into $N$ segments of length $\Delta L$ such that segment $i$ has position $\bm{Y}_i$ and frame $\{\bm{\hat{t}}_i,\bm{\hat{\mu}}_i,\bm{\hat{\nu}}_i\}$, as sketched in Figure \ref{fig:FF_schematic}B. After applying central differencing to \eqref{eq:continuous_equations3}, \eqref{eq:continuous_equations1} and \eqref{eq:continuous_equations2} and multiplying by $\Delta L$, we obtain the force and moment balance for each segment $i$,
\begin{align}
    \bm{F}_i^{C}+\bm{F}_i^H &=0,\label{eq:forcebal_discrete} \\
    \bm{T}_i^{E}+\bm{T}_i^{C}+\bm{T}_i^H &=0 \label{eq:torquebal_discrete},
\end{align}
and the discrete version of the kinematic constraint,
\begin{equation}
    \bm{Y}_{i+1}-\bm{Y}_i-\frac{\Delta L}{2}\left(\bm{\hat{t}}_i+\bm{\hat{t}}_{i+1}\right) =0. \label{eq:inextensibility_discrete}
\end{equation}
Here, $\bm{T}_i^{E}=\bm{M}_{i+1 / 2}-\bm{M}_{i-1 / 2}$ is the elastic torque, $\bm{F}_i^{C}=\bm{\Lambda}_{ i+1 / 2} - \bm{\Lambda}_{i-1 / 2}$ and $\bm{T}_i^C=(\Delta L/2) \bm{\hat{t}}_i \times (\bm{\Lambda}_{i+1 / 2}+\bm{\Lambda}_{ i-1 / 2})$ are the constraint forces and torques respectively, and $\bm{F}_i^H=\bm{f}_i^H\Delta L$ and $\bm{T}_i^H=\bm{\tau}_i^H \Delta L$ are the hydrodynamic force and torque on segment $i$. The internal moments, $\bm{M}_{i+1/2}$, are provided by a discrete version of the consitutive law, and $\bm{\Lambda}_{i+1/2},$ enforces (\ref{eq:inextensibility_discrete}).

Due to the negligible effects of fluid inertia, the fluid velocity is governed by the Stokes equations, and as a result, the forces and torques on the segments will be linearly related to their velocities and angular velocities via the mobility matrix, $\mathcal{M}$. Explicitly, we will have
\begin{equation}\label{eq:mobility_matrix_OG}
    \left(\begin{array}{l} \bm{V} \\ \bm{\Omega}\end{array}\right)= \mathcal{M} \left(\begin{array}{l}-\bm{F}^{H} \\ -\bm{T}^{H}\end{array}\right),
\end{equation}
where $\bm{V}$ and $\bm{\Omega}$ are, respectively, $3N \times 1$ vectors containing the components of the translational and angular velocities of all segments, and $\bm{F}^H$ and $\bm{T}^H$ are the corresponding $3N \times 1$ vectors containing the hydrodynamic forces and torques on all segments and given by Eqs (\ref{eq:forcebal_discrete}) and (\ref{eq:torquebal_discrete}).  For the $6N \times 6N$ mobility matrix, $\mathcal{M}$, we employ the Rotne-Prager-Yamakawa (RPY) tensor \cite{Wajnryb2013GeneralizationTensors} that is adapted to include the no-slip condition at $z=0$ \cite{Swan2007SimulationBoundary}. The explicit form that we use can be found in Appendix \ref{appen:RPY}. We set the hydrodynamic radius in the RPY tensor to be $a$, the cross-sectional radius of the filament.  We ensure also that $\Delta L$ is comparable to $2a$, as sketched in Figure \ref{fig:FF_schematic}B.

After solving the mobility problem for the velocities and angular velocities, we integrate the system forward in time and update the segment positions and frame vectors subject to the kinematic constraints.  Since filament deformation is fully 3D and so can involve both bending and twisting, we employ unit quaternions, $\bm{q} = (q_0, q_1, q_2, q_3)$ with $\lVert \bm{q} \rVert = 1$ \cite{Schoeller2021MethodsFilaments,Hanson2006VisualizingQuaternions}, to track the rotations of the frame vectors.  The unit-quaternion for segment $i$, $\bm{q}_i(t)$, maps the standard basis, $(\hat{\bm{x}},\hat{\bm{y}},\hat{\bm{z}})$, to the segment's frame vectors at time $t,$ $(\hat{\bm{t}}_i(t),\hat{\bm{\mu}}_i(t),\hat{\bm{\nu}}_i(t))$, through $\bm{R}(\bm{q}_i(t)) = (\hat{\bm{t}}_i(t) \; \bm{\hat{\mu}}_i(t) \; \bm{\hat{\nu}}_i(t))$, where $\bm{R}(\bm{q})$ is the rotation matrix,
\begin{equation}
    \bm{R}(\bm{q}) = 
    \begin{pmatrix}
    1-2q_2^2-2q_3^2 \; & \; 2(q_1 q_2 - q_3q_0) \; & \; 2(q_1q_3 + q_2q_0) \\
    2(q_1q_2+q_3q_0) \; & \; 1-2q_1^2-2q_3^2 \; & \; 2(q_3q_2-q_1q_0) \\
    2(q_1q_3 - q_2q_0) \; & \; 2(q_3q_2+q_1q_0) \; & \; 1-2q_2^2 - 2q_1^2
    \end{pmatrix}.
\end{equation}

The segment positions and quaternions evolve according to the differential-algebraic system 
\begin{align}
    \frac{d \bm{Y}_i}{d t} &=\bm{V}_i, \label{eq:continuous_equation_timeevo1} \\
    \frac{d \bm{q}_i}{dt} &= \frac{1}{2} (0, \bm{\Omega}_i) \bullet \bm{q}_i,\label{eq:continuous_equation_timeevo2}\\ 
    \bm{Y}_{i+1}-\bm{Y}_i-\frac{\Delta L}{2}\left(\bm{\hat{t}}_i+\bm{\hat{t}}_{i+1}\right) &=0,
\end{align}
for each $i$, where $\bullet$ is the quaternion product, defined for two quaternions $\bm{p} = (p_0, \tilde{\bm{p}})$ and $\bm{q} = (q_0, \tilde{\bm{q}})$ as $\bm{p} \bullet \bm{q} = (p_0 q_0 - \tilde{\bm{p}}\cdot \tilde{\bm{q}}, \; p_0\tilde{\bm{q}} + q_0 \tilde{\bm{p}} + \tilde{\bm{p}} \times \tilde{\bm{q}})$.  We discretise these equations in time using a second-order, geometric BDF scheme described in \cite{Schoeller2021MethodsFilaments}, which preserves the unit-length of the quaternions as they are updated.  The resulting equations along with the kinematic constraints given by \eqref{eq:inextensibility_discrete} provide a nonlinear system that we then solve iteratively using Broyden's method \cite{Broyden1965AEquations}, again following the scheme presented in \cite{Schoeller2021MethodsFilaments}.

\subsection{Simulation Parameters}
In this paper, the segment length is $\Delta L = 2.2a$ and the bending modulus is taken to be equal to the twist modulus, $K_B=K_T$. We justify this choice, and discuss the effect of varying the twist modulus on our results, in Appendix \ref{appen:twist}. We investigate the effect of varying two non-dimensional parameters; the nondimensional follower force, $f,$ as defined in Eq (\ref{eq:followerforce_BC}), and the aspect ratio, $\alpha = L/a$. The aspect ratio is adjusted while keeping the bending modulus fixed, corresponding to varying the filament length.  This results in changing the hydrodynamic response to the filament, while keeping its elastic response fixed. All timescales are reported with respect to the filament's relaxation time, $\tau$, the characteristic timescale associated with a bent filament returning to its undeformed configuration. We note that the relaxation time varies with the aspect ratio. We obtain this timescale numerically as described in Appendix \ref{appen:relaxation_timescale}.

%% file: part3.tex
\section{State space}\label{sec:ivp}

\begin{figure}
    \centering
    \includegraphics[width=0.75\textwidth]{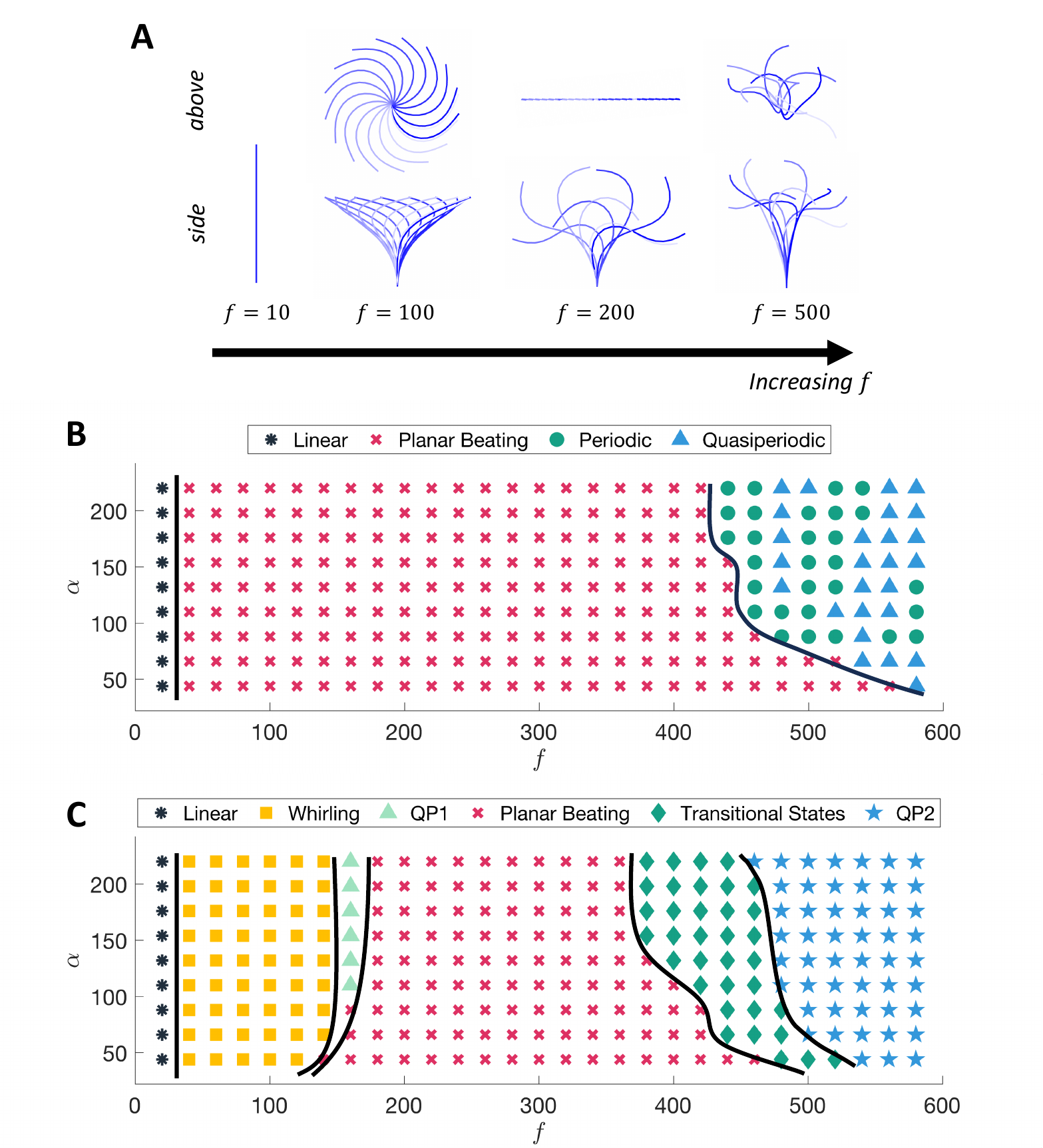}
    \caption{A: Images showing the steady, whirling, planar beating and writhing states for $\alpha = 44$ (from left to right). Views of the $(x,z)-$ plane (side) and $(x,y)-$plane (above) are provided and the passage of time is indicated by the increasing boldness of the lines. B: The state space of solutions over a range of $\alpha$ and $f$  when dynamics are restricted to 2D.  C: The state space of solutions over a range of $\alpha$ and $f$ for fully 3D dynamics. In panels B and C, the black lines are drawn to indicate the approximate boundaries of the solutions regions.}
    \label{fig:picturesosfar}
\end{figure}

Before performing a detailed bifurcation and stability analysis, we first solve a series of initial value problems using the methods outlined in Section \ref{sec:ActiveFilamentModel} to provide a general overview of the $(f,\alpha)$ state space and demonstrate that our code reproduces the results found in previous studies.  Unless otherwise stated, in each simulation the filament is initially straight and untwisted, i.e. $(\hat{\bm{t}}_i(0),\hat{\bm{\mu}}_i(0),\hat{\bm{\nu}}_i(0)) =(\hat{\bm{e}}_z,\hat{\bm{e}}_x,\hat{\bm{e}}_y) $ for all $i$, and a perturbation is introduced in the form of a small, random force applied to three of the segments at the first timestep. If the random forces are co-planar, the resulting dynamics will be in that plane, otherwise, the dynamics can be fully 3D.

With the filament aspect ratio fixed at $\alpha = 44$, we obtain results consistent with those from previous studies of planar \cite{DeCanio2017} and fully 3D \cite{Ling2018Instability-drivenMicrofilaments} filament motion.  For planar perturbations, we observe that below a critical follower force value, $f^*$, the perturbation decays with time and the filament returns to its vertical equilibrium. Above the critical value, however, we observe the onset of self-sustained symmetric oscillations which we call planar beating. When motion is fully 3D, we observe buckling at the same critical value, $f^*$, but instead of planar beating, a fully 3D whirling motion emerges.  Whirling can be characterised as a rigid body rotation of the filament centerline, with the filament tip tracing out a circle in the $(x,y)-$plane. Increasing the forcing, we find evidence of a second bifurcation as the filament instead performs the same planar beating observed in the 2D model.  At very high force values, filament motion enters a writhing regime where it undergoes non-periodic oscillations, frequently changing direction and beat shape. These four distinct behaviours are depicted in Figure \ref{fig:picturesosfar}A.

Exploring the state space further for higher values of $\alpha$, we find that new solutions emerge.  The state space when dynamics is restricted to 2D is shown in Figure \ref{fig:picturesosfar}B.  Here, we see that for $f > 400$, planar beating transitions to a regime where one encounters a mix of quasiperiodic solutions and time-periodic solutions more complicated than planar beating.  The force value where this transition occurs decreases as $\alpha$ increases.  The state space for fully 3D dynamics is shown in Figure \ref{fig:picturesosfar}C.  Here, we find the existence of a new, quasiperiodic solution (QP1) that emerges for a small range of $f$ between the whirling and the beating solutions.  Furthermore, we see that the region broadly described as writhing can be further divided into two distinct domains of behaviour. The first is a transition region where beating becomes unstable and gives rise to a diverse array of behaviours highly dependent on $\alpha$.  In the second, occurring at higher $f$, we find a second quasiperiodic solution (QP2 in Figure \ref{fig:picturesosfar}C). 

In the subsequent sections, we explore these solutions and study the bifurcations that produce them.  A limitation of solving initial value problems is that they only reveal behaviours that are stable.  In studying the solutions and bifurcations, we move beyond initial value problems by using a Jacobian-Free Newton Krylov (JFNK) method to track the time-periodic solutions of beating and whirling into regions where they are unstable and perform a linear stability analysis to classify observed bifurcations.  We also analyse the new quasiperiodic solutions in depth, using Poincar\'{e} sections to elucidate their time-dependence and further illustrate how the solution space changes with filament aspect ratio.
%

%% file: part4.tex



\section{Jacobian-free Newton Krylov method and stability analysis}
\subsection{Computation of time-periodic solutions}\label{Section:bifurcation_theory}

To understand the bifurcations yielding the different states, we need a method to track solutions on either side of the bifurcation.  We can track time-periodic solutions, even at forcing values where they unstable, using an adapted Newton-Raphson approach described in \cite{Viswanath2007RecurrentTurbulence,Willis2019EquilibriaThem}.  For the state variable that describes the filament configuration at any given time, we utilise the Lie algebra elements, $\bm{v}_1,...,\bm{v}_N$, that generate the quaternions of each segment with respect to the vertical equilibrium.  Accordingly, for each segment $i$, $\bm{q}_i$ is related to $\bm{v}_i$ through the expression,
\begin{equation}\label{eq:effective_lie_alg}
    \bm{q}_i = \exp(\bm{v}_i) \bullet \bm{q}_{VE},
\end{equation}
where $\bm{q}_{VE} = \frac{1}{\sqrt{2}} [1,0,1,0]$ is the quaternion corresponding to a vertically upright filament, and the exponential map is 
\begin{equation}
   \exp(\bm{u}) = \left(\cos\left(\frac{||\bm{u}||}{2}\right), \sin\left(\frac{||\bm{u}||}{2}\right)\frac{\bm{u}}{||\bm{u}||} \right).
\end{equation}
Thus, from $\bm{q}_i$, we can extract the corresponding Lie algebra element $\bm{v}_i = ||\bm{v}_i|| \hat{\bm{v}}_i,$ through
\begin{align}
    \hat{\bm{v}}_i &= \frac{[\bm{q}_i \bullet \bm{q}_{VE}^{*}]_{\mathbb{R}^3}}{||[\bm{q}_i \bullet \bm{q}_{VE}^{*}]_{\mathbb{R}^3}||}, \label{eq:effective_lie_alg_from_q_1}\\
    ||\bm{v}_i|| &= 2\arccos\left((\bm{q}_i \bullet \bm{q}_{VE}^{*}) \cdot \hat{\bm{e}}_1\right), \label{eq:effective_lie_alg_from_q_2}
\end{align}
where $\hat{\bm{e}}_1 = [1,0,0,0]$ and $[\bm{q}]_{\mathbb{R}^3} = (q_1, q_2, q_3)$.  

In this set-up, the trivial state, $\bm{v}_1=...=\bm{v}_N = \bm{0}$, corresponds to the steady vertically upright filament.  We note also that due to the clamped end condition, we have $\bm{v}_1(t)=\bm{0}$ for all time.  As a result, any admissible filament configuration can be reached using the state variable $\bm{U}(t) = (\bm{v}_2(t),...,\bm{v}_N(t))$.

With the state variable in hand, to implement the JFNK method from \cite{Viswanath2007RecurrentTurbulence,Willis2019EquilibriaThem,Willis2017TheSolver}, we must now define how it evolves in time.  At the temporally continuous level, the time evolution can be summarised as follows.  We first transform the $\bm{v}_i(t)$ contained in $\bm{U}(t)$ into their corresponding $\bm{q}_i(t)$ via Eq (\ref{eq:effective_lie_alg}).  From the quaternions, we determine the filament configuration at time $t$ and solve \eqref{eq:inextensibility_discrete}, \eqref{eq:continuous_equation_timeevo1} and \eqref{eq:continuous_equation_timeevo2} to advance the $\bm{q}_i(t)$ to time $t+t'$.  Finally, we compute $\bm{v}_i(t + t')$, and hence $\bm{U}(t+t')$ using \eqref{eq:effective_lie_alg_from_q_1} and \eqref{eq:effective_lie_alg_from_q_2}.  This process corresponds to applying the continuous flow-map, $\bm{\varphi}(\bm{U},t),$ which describes the evolution of the state variable over an interval of time. Specifically, we have
\begin{equation}\label{eq:varphi_definition}
    \bm{U}(t + t') = \bm{\varphi} (\bm{U}(t),t'),
\end{equation}
for all times $t$, $t'$. 

Using the flow map, time-periodic solutions are then solutions to
\begin{equation}\label{eq:Willis_nonlinearsystem}
    \bm{G}(\bm{U},T) = \bm{\varphi}(\bm{U},T) - \bm{U} = \bm{0}
\end{equation}
for some nonzero period, $T$. The nonlinear system \eqref{eq:Willis_nonlinearsystem} for $\bm{U}$ and $T$ provides $3N-3$ equations and is therefore underdetermined.  An additional equation is necessary and following \cite{Viswanath2007RecurrentTurbulence, Viswanath2009TheNumber,Willis2019EquilibriaThem,Willis2017TheSolver}, we will include it as part of the Newton iteration.



For this purpose, we define a new variable $\bm{X} = (\bm{U},T)$.  We denote the initial guess for the solution as $\bm{X}_0=(\bm{U}_0,T_0)$ and seek to determine its update $\bm{X}_0 + \tilde{\bm{X}}$, where $\tilde{\bm{X}} = (\tilde{\bm{U}},\tilde{T})$. Linearising \eqref{eq:Willis_nonlinearsystem} about the initial guess yields 
\begin{equation}\label{eq:Newton_variation}
\bm{G}(\bm{X}_0) + \bm{J}_{\bm{G}}(\bm{X}_0) \tilde{\bm{X}} = \bm{0},
\end{equation}
where
\begin{equation}\label{eq:Jacobian_definition}
    \bm{J}_{\bm{G}}(\bm{X}_0) = \left. \frac{\partial \bm{G}}{\partial \bm{X}} \right|_{\bm{X}_0}
\end{equation}
is the Jacobian of $\bm{G}(\bm{X})$ evaluated at $\bm{X}_0$.  Following \cite{Viswanath2007RecurrentTurbulence}, we incorporate the additional equation by insisting that $\tilde{\bm{U}}$ is perpendicular to its velocity at $T$, preventing any update of the solution along its periodic orbit.  Specifically, the additional equation is
\begin{equation}\label{eq:Newton_constraint}
    \bm{\tilde{U}} \cdot \dot{\bm{U}_0} = 0,
\end{equation}
where $\dot{} = {d}/{dt}$. The equations \eqref{eq:Newton_variation} and \eqref{eq:Newton_constraint} combine to provide the linear system, 
\begin{equation}
    \bm{A} \bm{\tilde{X}} = \bm{b}
\end{equation}
that we need to determine $\tilde{\bm{X}}$ at iteration $k$, where
\begin{equation}
    \bm{A} \tilde{\bm{X}}= \begin{bmatrix}
        \bm{J}_{\bm{G}}(\bm{X}_k) \tilde{\bm{X}} \\ 
        \dot{\bm{U}}_k \cdot \tilde{\bm{U}}
    \end{bmatrix},
\end{equation}
and
\begin{equation}
    \bm{b} = \begin{bmatrix}
        -\bm{G}(\bm{X}_k) \\ 0
    \end{bmatrix}.
\end{equation}
We solve this linear system using the Generalised Residual Method (GMRES). The application of the Jacobian is done using the matrix-free difference formula,
\begin{equation}\label{eq:Willis_Jacobianapprox}
    \bm{J}_{\bm{G}}(\bm{X}_k) \bm{\tilde{X}} = \frac{1}{\epsilon} \left( \bm{G}(\bm{X}_k + \epsilon \bm{\tilde{X}}) - \bm{G}(\bm{X}_k) \right) + O(\epsilon),
\end{equation}
where $\epsilon \ll 1$ is a small parameter and $\dot{\bm{U}}_k$ is evaluated using a first-order forward difference. Once GMRES converges, the solution is updated via $\bm{X}_{k+1} = \bm{X}_k + \bm{\tilde{X}}$, which we accelerate further using a locally constrained optimal hook step as described in \cite{Viswanath2007RecurrentTurbulence,Viswanath2009TheNumber}.  Once the Newton method converges and we obtain a time-periodic solution, we use continuation to track the solution for different values of $f$, including values of $f$ for which the solution may be unstable.

\subsection{Linear stability analysis}\label{sec:LinearStabilityAnalysisTheory}
Along with finding time-periodic solutions, we analyse their stability, as well as the stability of the trivial steady state. 
To do so, we first express the continuous-time system given by (\ref{eq:inextensibility_discrete}), (\ref{eq:continuous_equation_timeevo1}) and (\ref{eq:continuous_equation_timeevo2}) for the state variable as
\begin{equation}\label{eq:continuoussystem_stability}
    \frac{d\bm{U}(t)}{dt} = \bm{f}(\bm{U}(t)).
\end{equation}
Consider a small perturbation $\delta\bm{U}$ about a base state $\bm{U}_0$, such that $\bm{U}=\bm{U}_0+\delta\bm{U}$. The linearised dynamics for the small perturbation from \eqref{eq:continuoussystem_stability} is given by
\begin{equation}\label{eq:continuoussystem_stability2}
 \frac{d\delta\bm{U}(t)}{dt} =  \bm{A}\delta\bm{U},
\end{equation}
where $\bm{A}=\partial\bm{f}/\partial\bm{U}|_{\bm{U}_0}$. The general solution to (\ref{eq:continuoussystem_stability2}) is given by $\delta\bm{U}(t)=\bm{\Phi}(t;\bm{U}_0)\delta \bm{U}(0)$, where $\Phi(\bm{U}_0)$ is the state transition operator (or impulse response).  When $\bm{U}_0$ is a steady solution, 
\begin{equation}\label{eq:phi_steady}
\bm{\Phi}(t;\bm{U}_0)=e^{t\bm{A}}.
\end{equation}
When $\bm{U}_0$ is time-periodic with period $T$, i.e. $\bm{U}_0(t)=\bm{U}_0(t+T)$, Floquet's theorem gives
\begin{equation}\label{eq:phi_timeperiodic}
\bm{\Phi}(t;\bm{U}_0)=\bm{P}(t)e^{t\bm{B}},
\end{equation}
where $\bm{P}(t)=\bm{P}(t+T)$ with $\bm{P}(0)=\bm{I}$ and $\bm{B}$ is the Floquet matrix exponent. 

The stability of steady and time-periodic solutions in this study is examined by computing the eigenvalues of $\bm{\Phi}(T;\bm{U}_0)$, where $T$ is an arbitrary short time interval for a steady solution, or the time period for a periodic solution. For this purpose, time integration is combined with the Arnoldi method to extract the approximate eigenvalues of $\bm{\Phi}(T;\bm{U}_0)$ (say $\mu$) by advancing the initial condition $\bm{U}_0(0) + \delta\bm{U}(0)$ to time $T$.  From $\mu$, the eigenvalues, $\lambda$, of $\bm{A}$ or $\bm{B}$ can then be determined from
\begin{equation}
 \lambda = \frac{1}{T} \log (\mu).
\end{equation}
We provide further details of this method in Appendix \ref{appen:arnoldi}.

%% file: part5.tex
\section{Double Hopf bifurcation}\label{sec:doublehopf}


\begin{figure}
    \centering
    \includegraphics[width = \textwidth]{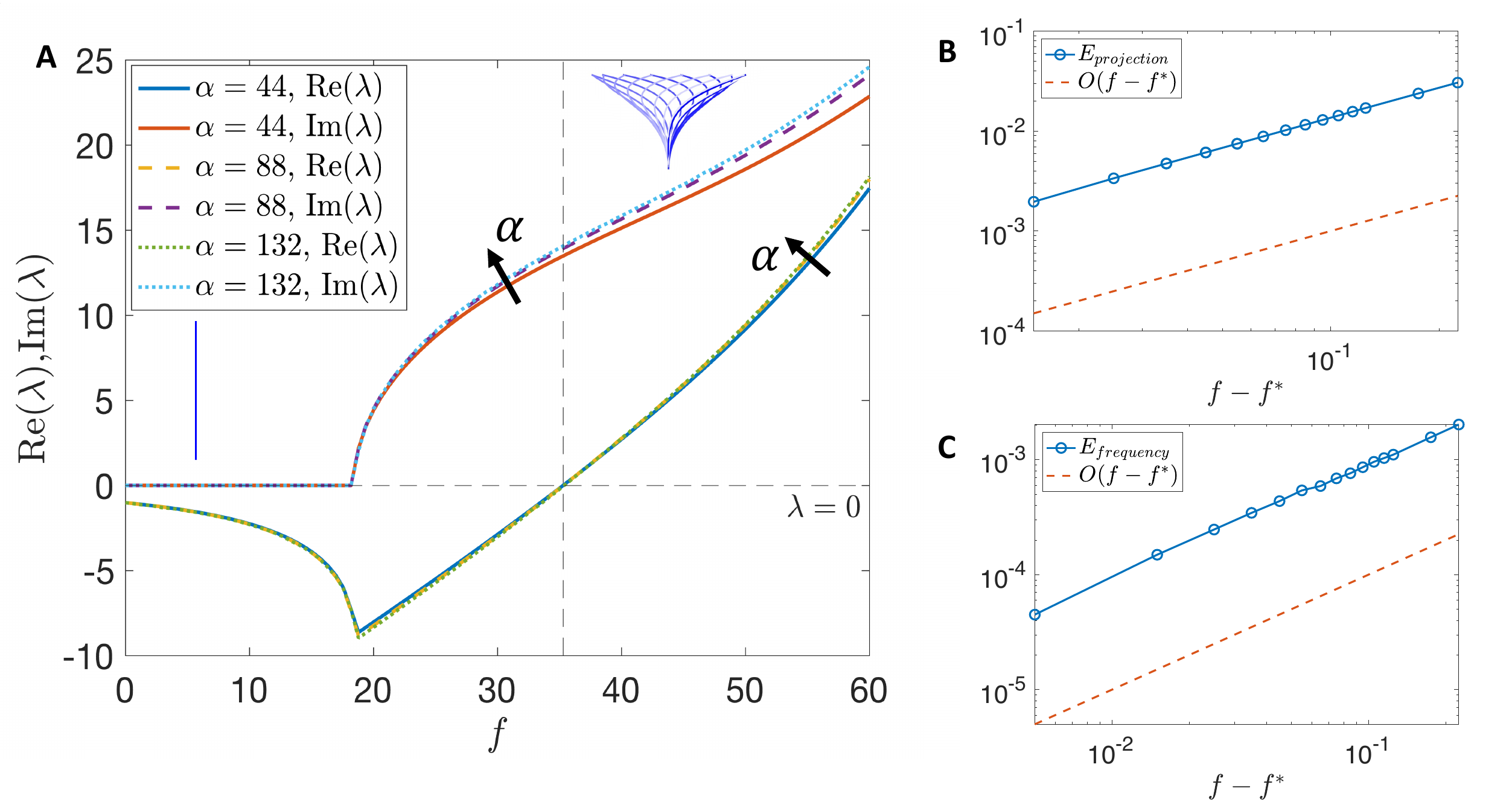}
    \caption{A: The dominant eigenvalue from the linear stability analysis of the trivial steady state for $f \in [0,60]$ and three values of $\alpha$. The steady state becomes unstable at $f^* \approx 35.3$ regardless of the aspect ratio.  The eigenvalues are purely imaginary at the bifurcation. B: The difference between the solution and its the projection onto the eigenspace at the bifurcation, $E_{\text{projection}}:=\text{min}_{a_x,a_y \in \mathbb{C}}||\bm{U} - Re(a_x \bm{\zeta_x}) - Re(a_y \bm{\zeta_y}) ||,$ grows like $f-f^*$ close to the bifurcation. C: The difference between the whirling frequency and the imaginary part of the eigenvalue at the bifurcation, $E_{\text{frequency}} := ||2\pi /T - \omega ||$, for $\alpha = 44.$ The difference grows like $f-f^*$ near the bifurcation.}
    \label{fig:doublehopf}
\end{figure}

By solving the initial value problem, we have seen that above a critical forcing, $f^*,$ the filament buckles. In 2D buckling leads to beating, while in fully 3D simulations whirling is observed. To reconcile these observations, we perform a linear stability analysis of the trivial state (vertically upright filament) using the Arnoldi method described in Section \ref{sec:LinearStabilityAnalysisTheory}. This computation yields the dominant eigenmodes which correspond to the least stable states.  We extract the dominant eigenvalue, i.e. the eigenvalue with the largest real part and hence the fastest growing mode, and its associated eigenvector. The eigenvector provides the filament shape that will grow in amplitude with time. 

Performing this analysis for different $\alpha$, as shown in Figure \ref{fig:doublehopf}A, indicates modest increases in the real part of the dominant eigenvalue as $\alpha$ increases.  We note, however, that the real part of the eigenvalue becomes positive at $f^* \approx 35.3$ for each $\alpha$, demonstrating that the critical value for buckling is independent of $\alpha.$ The imaginary part of the eigenvalue that provides the frequency of the unstable solution near the bifurcation increases slightly with the aspect ratio.

To connect the differences in the 2D and 3D solutions at the bifurcation, we first find the 2D unstable modes using a perturbation $\delta \bm{U}(0)$ that restricts motion to the $(x,z)$-plane.  At the bifurcation, the dominant eigenvalues are purely imaginary, and so can be expressed as $\lambda_{\pm} = \pm i \omega.$ Their associated eigenvectors are the complex conjugate pair, say $\bm{\zeta}_x, \bar{\bm{\zeta}}_x$.  We note that by rotational symmetry, an orthogonal set of eigenvectors, $\bm{\zeta}_y,$ $\bar{\bm{\zeta}}_y$, which correspond to beating in the transverse plane, exist with the same eigenvalues.

Turning our attention now to the full 3D problem by removing any restrictions on $\delta \bm{U}(0)$, we find that the dominant eigenvalues are identical to those found in 2D, $ \lambda_{\pm} = \pm i \omega$. However, we now find that there is not only one, but two, complex conjugate pairs of unstable eigenvectors.  While we do find that the eigenvectors depend on the initial perturbation used to compute them, they can always be written as linear combinations of the two sets of planar eigenvectors. Hence we can express the unstable eigenmodes at the bifurcation as
\begin{align}\label{eq:eigen}
\begin{split}
    (i\omega,\bm{\zeta}_x), (-i\omega,\bar{\bm{\zeta}}_x), \\ (i\omega,\bm{\zeta}_y), (-i\omega,\bar{\bm{\zeta}}_y).
\end{split}
\end{align} 

A bifurcation with two pairs of imaginary eigenvalues is defined as a double Hopf, or Hopf-Hopf, bifurcation \cite{kuznetsov1998elements}, as opposed to a Hopf bifurcation where only a single pair is present.  While in general for a double Hopf bifurcation the eigenvalue pairs can be different, in our specific problem, we obtain repeated pairs due to the rotational symmetry that is present.  Bifurcations in the presence of symmetry have been analysed previously using group theoretic techniques \cite{golubitsky1985hopf}.  We also note that a double Hopf bifurcation occurs in systems with two parameters.  The fact that we encounter a double Hopf bifurcation here while varying only the follower force suggests the presence of another hidden parameter already at its critical value. Variation of the hidden parameter must break the rotational symmetry of the two unstable modes, or in other words, the isotropy of unstable motion in plane normal to the filament tangent. Given that the instability is related to buckling, such a hidden parameter may be bending modulus anisotropy, for example.  Anisotropy would then allow buckling to occur in one direction at a lower follower force value compared to the other, resulting in Hopf bifurcations at different critical values of the follower force.  Only when the bending moduli in both directions are equal would we see the double Hopf bifurcation.


To better understand the system near the bifurcation, we propose a form of the solution just after it occurs following standard weakly nonlinear analysis \cite{Stuart1960,Kuramoto1983ChemicalTurbulence.,Schmid01} (for the details in the context of fluid mechanics, see Eq. (5.48) in \cite{Schmid01}). The solution near the bifurcation is spanned by the unstable eigenvectors with amplitudes that vary over a slow time scale associated with the small growth rate of the instability. When the amplitudes have steady values, as in the beating and whirling states in this study, the solution may be written as
\begin{equation}\label{eq:42}
    \bm{U} \approx \bm{U}^* + A_1 \sqrt{f-f^*} \bm{\zeta}_x e^{i \omega t} + A_2 \sqrt{f-f^*} \bm{\zeta}_y e^{i\omega t} + \text{c.c.} + O(f-f^*),
\end{equation}
for $|f-f^*|\ll 1$, where $\bm{U}^*$ is the unstable steady state solution and $A_1$ and $A_2$ are constant complex amplitudes.  Indeed, we find that the error in the projection of the numerical whirling solution onto the eigenspace scales like $\sim f-f^*$, as shown in Figure \ref{fig:doublehopf}B.  Eq. (\ref{eq:42}) is also consistent with the observation that the growth of the solution scales like $\sqrt{f-f^*}$ (not shown) and that the error in the frequency scales like $f-f^*$ (see Figure \ref{fig:doublehopf}C).  We note that for a standard Hopf bifurcation the solution would only contain a single mode and its associated amplitude (along with its complex conjugate).  Here, for the double Hopf bifurcation that we encounter, we have two modes.  The additional dimensions in which solutions live allow for both beating and whirling to emerge at a single bifurcation.
From the rotational symmetry of the basic state (i.e. the straight filament), once the filament is saturated into the whirling state we have $|A_1|=|A_2|$ as confirmed by our numerical simulation. The precise values of the complex amplitudes are, however, to be determined from the initial conditions as well as from nonlinear interactions between the two modes. 

The double Hopf bifurcation with a solution of the form \eqref{eq:42} implies that for the 2D case, either $A_1$ or $A_2$ must be zero and the filament would beat in the corresponding plane with angular frequency $\omega$. Based on the stability analysis, we expect beating to be unstable to orthogonal perturbations. However, in 3D, the filament may execute rotational motion about $z$-axis, as both $A_1$ and $A_2$ are expected to be non-zero and incorporate the phase difference needed to achieve the whirling state. Saturation to any state would require a nonlinear mechanism to balance the growth in the unstable modes, however the lack of preferential direction in the problem suggest that solutions that do emerge, in this case circular whirling, must reflect this symmetry.  Since the eigenvectors are geometrically orthogonal and their size may be normalised arbitrarily, suppose the location of the filament tip in $(x,y)$-plane is given by two sets of eigenvectors, $(1,0)$ and $(0,1)$, respectively. Then, \eqref{eq:42} implies that that tip location in time would be
\begin{equation}\label{eq:xy}
    (x,y)_{\mathrm{tip}} \approx (B_1,B_2) \sqrt{f-f^*} e^{i \omega t} + \text{c.c.},
\end{equation}
where $B_1$ and $B_2$ are complex constants with $|B_1|=|B_2|$. Eq. \eqref{eq:xy} indicates that the tip location is, in general, an ellipse with its characteristic angular frequency $\omega$. The whirling state observed in Section \ref{sec:ivp} is admitted when $B_1$ and $B_2$ have a particular phase difference: $|\mathrm{Arg}(B_1)-\mathrm{Arg}(B_2)|=\pi/2$. This suggests that there is a nonlinear interaction between the two linear instability modes in (\ref{eq:42}) (e.g. resonance) to form the rotationally invariant whirling state.  

The whirling state can be viewed as a steady state in a rotating frame where separate components of the follower force are balanced by the other forces that are present.  The vertical (surface normal) component of the follower force is resolved mostly by the internal stresses in the filament, resulting in the filament's nontrivial centreline shape without changing in time. The horizontal component, however, is instead balanced mostly by the viscous stresses from fluid on the filament, which may further deform the filament, but more importantly establishes the constant angular velocity associated with the whirling solution. In other words, for the whirling state, the follower force is weak enough, such that its horizontal component does not require large internal stresses to achieve force and moment balance, resulting in a rotational motion with a constant angular velocity. However, if the follower force is increased further, the viscous stresses alone are not able to resolve it. In this case, relatively large amounts of internal stresses along the filament would be required to form a balance with the follower force. This presumably generates an instability, which ultimately leads to an unsteady motion with larger bending along the filament, i.e. the beating state, as we discuss below.

%% file: part6.tex

\section{From whirling to beating}\label{sec:whirlingtobeating}

\begin{figure}
    \centering
    \includegraphics[width=\textwidth]{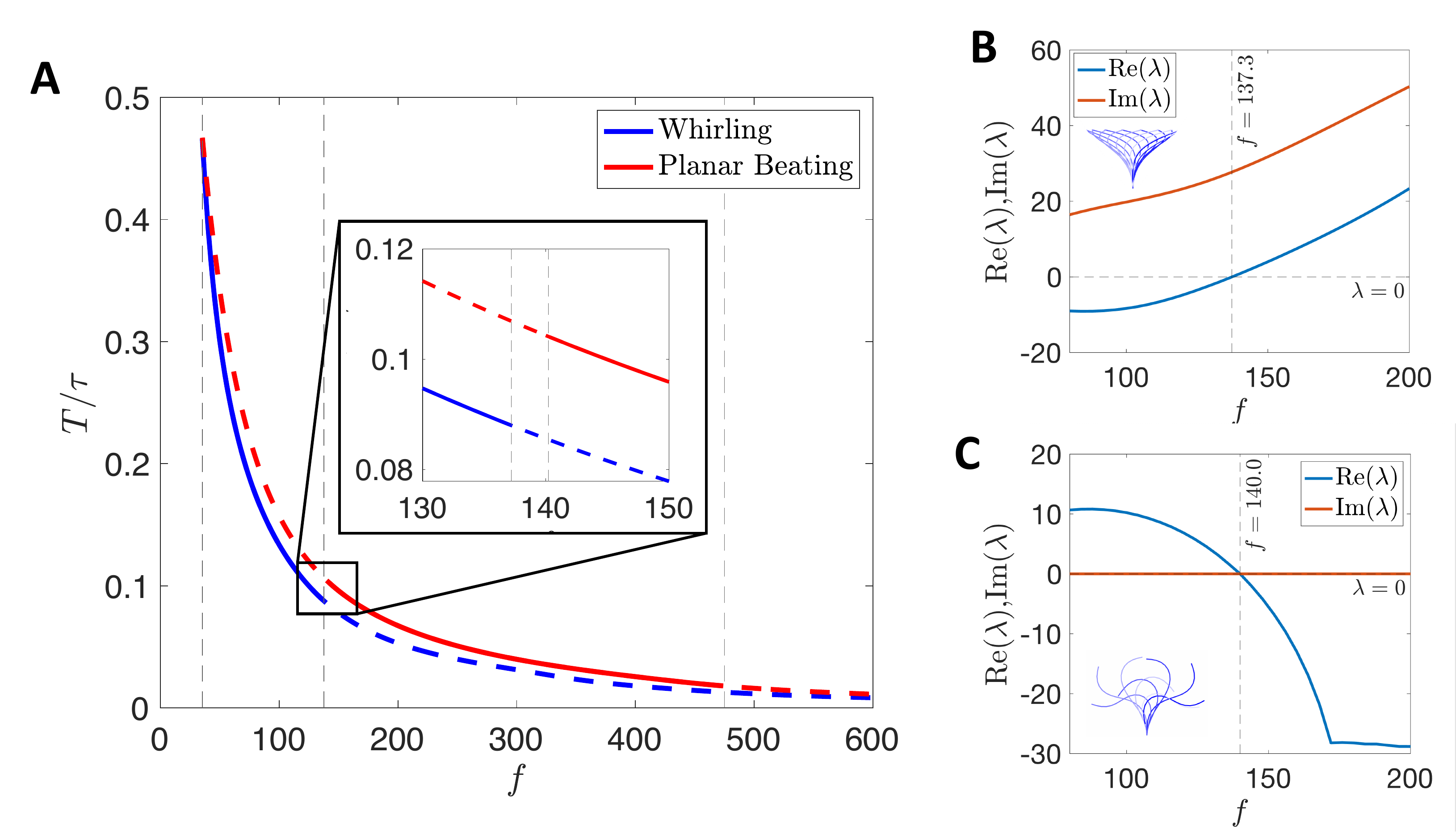}
    \caption{A: The period of the whirling (blue) and planar beating (red) solutions for $f \in [f^*,600]$ and $\alpha = 44.$ Solid (dashed) lines correspond to stable (unstable) solutions. Inset: A close up showing the range $f\in[130,150]$ indicating that two bifurcations occur. B: Eigenvalues from the Floquet analysis of the whirling solution for $f\in[80,200]$. Whirling becomes unstable at $f=137.3.$ C: Eigenvalues from the Floquet analysis of the beating solution for $f\in[80,200]$. Beating becomes stable at $f=140.0.$}
    \label{fig:secondbifurcation}
\end{figure}

We now consider the next bifurcation in the 3D problem, where the whirling state becomes unstable. Previous work \cite{Ling2018Instability-drivenMicrofilaments,Westwood2021CoordinatedSurfaces} based on solving initial value problems points to a sharp transition between the whirling and beating behaviours at $f \approx 137.5$ for $\alpha=44$. In order to elucidate this bifurcation, we track the beating and whirling branches as $f$ increases using the JFNK approach described in Section \ref{Section:bifurcation_theory}. We track the whirling and beating branches for $f\in [f^*,600]$.  Figure \ref{fig:secondbifurcation}A shows how the period of these solutions varies with $f$ for $\alpha=44$. Changing $\alpha$ produces the same qualitative behaviour and the quantitative differences that arise are due differences in the bifurcation points, and that the period decreases with $\alpha$.  In performing this continuation, we see that the planar beating branch can be tracked back to the initial buckling.  We can assess the stability of these time-periodic solutions by computing their corresponding Floquet exponents for $f\in [f^*,600]$ using the methods from Section \ref{sec:LinearStabilityAnalysisTheory}.  The Floquet exponents shown in Figure \ref{fig:secondbifurcation} reveal that for $\alpha=44,$ whirling becomes unstable at $f = 137.3$ (see Figure \ref{fig:secondbifurcation}B), while beating becomes stable at $f = 140.0$ (see Figure \ref{fig:secondbifurcation}C). Thus there is a region of $f$ where both behaviours are unstable.

\begin{figure}
    \centering
    \includegraphics[width=\textwidth]{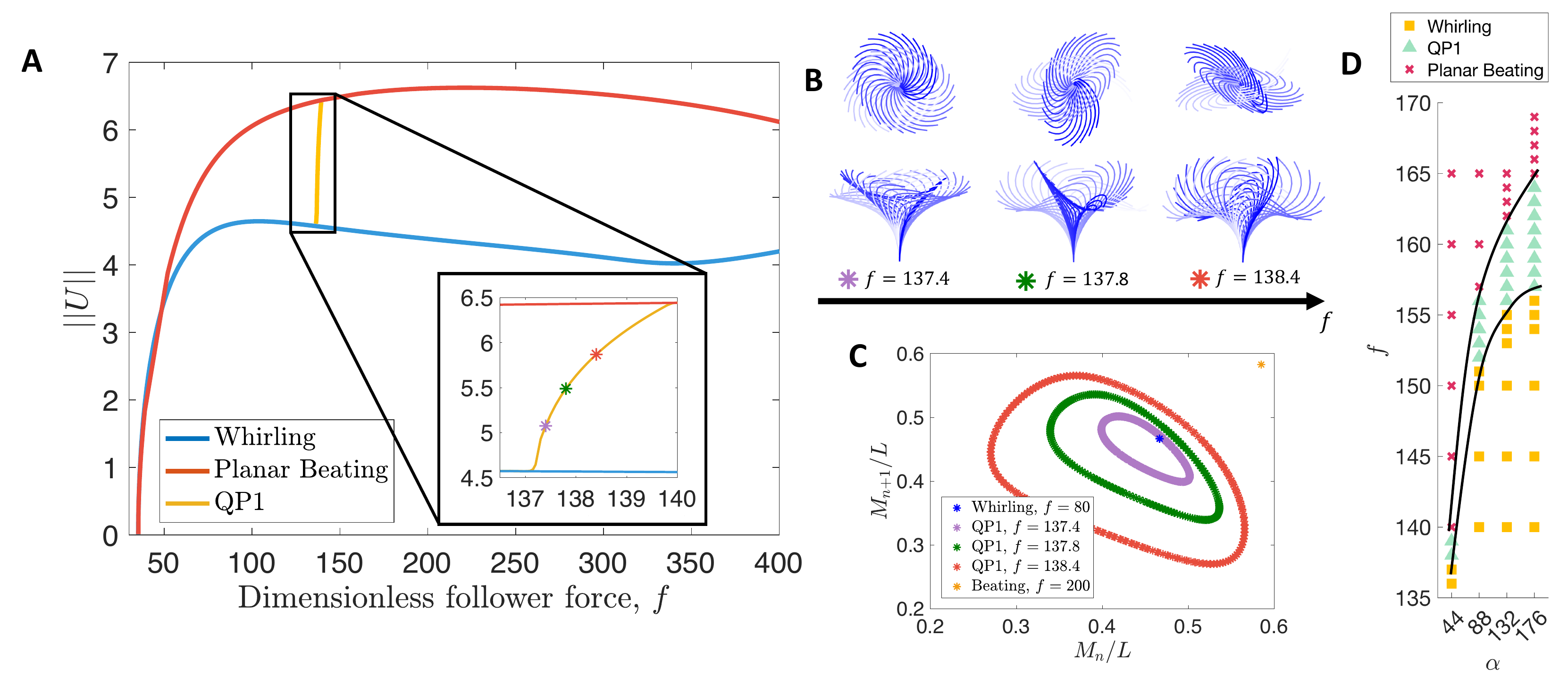}
    \caption{A: The maximum norm of the Lie algebra elements, $||\bm{U}||$, for the whirling, beating and quasiperiodic solutions as a function of $f.$ Inset: Close up of the  quasiperiodic branch connecting with the whirling and beating branches. The three markers indicate the solutions shown in B and C. B: Images of the filament in the quasiperiodic regime for three values of $f$.  The passage of time is indicated by an increasing in the boldness of the lines. The eccentricity of the ellipse traced by the filament tip decreases with increasing $f$.  C: A Poincar\'{e} section of the whirling and beating solutions and QP1 for different $f$. 
    D: The $(\alpha,f)$ state space in the region where QP1 exists.  The range of $f$ for which QP1 exists increases with $\alpha$.}
    \label{fig:quasiperiodicsolution}
\end{figure}

Indeed, there is a stable solution that connects the beating and whirling solutions (see Figure \ref{fig:quasiperiodicsolution}A) and thus there are two bifurcations.  As this solution is stable, we track this branch by solving initial value problems. We observe that this stable solution appears as a mixture of the whirling and beating.  Two distinct frequencies, one for beating and one for whirling, are associated with this solution, and, in the correct rotating frame, the filament tip traces out an ellipse with an eccentricity that increases continuously with $f$ (see Figure \ref{fig:quasiperiodicsolution}B).  The two periods are not integer divisors, implying that this solution is quasiperiodic. Accordingly, we refer to this solution as QP1.

To study QP1 in more detail, we construct Poincar\'{e} sections using a time series of the $x-$coordinate of the filament tip. We label successive local maxima of this signal as $M_1,M_2,...$, and then plot consecutive maxima against each other. We note that when the filament dynamics are planar, the plane of beating is arbitrary and hence, without loss of generality, we take the beat plane to be the $(x,z)-$plane.  
Time periodic solutions, such as whirling and beating, appear as single points on the Poincar\'{e} section, as shown in Figure \ref{fig:quasiperiodicsolution}C. Quasiperiodic solutions, on the other hand, appear as curves. In Figure \ref{fig:quasiperiodicsolution}C, we see that QP1 yield closed curves, consistent with its quasiperiodic behaviour and also characterising the solution as a 2-torus. As the forcing increases the diameter of the 2-torus trace increases due to the increasing eccentricity of the ellipse traced by the tip. 

Using Floquet analysis further, we can classify the two bifurcations that occur.  Figure \ref{fig:secondbifurcation}B reveals that the eigenvalue from the whirling branch is imaginary at the bifurcation and is distinct from the period of the whirling solution. Thus, the emergence of the quasiperiodic branch corresponds to a supercritical Hopf bifurcation. 
On the other hand, we observe that the eigenvalue from the planar beating branch is purely real (see Figure \ref{fig:secondbifurcation}C). This indicates that the second bifurcation is a pitchfork bifurcation, whereby the two stable solutions (corresponding to the quasiperiodic solution rotating clockwise or anticlockwise) cease to exist after beating becomes stable.
We investigate how $\alpha$ affects these bifurcations. Increasing $\alpha$ delays the onset of the first bifurcation and broadens the domain of $f$ for which QP1 exists (Figure \ref{fig:quasiperiodicsolution}D). 

While ascertaining the precise reasons for the onset of QP1 requires a more detailed analysis, we can gain some understanding of this bifurcation by examining the unstable eigenmode just after the whirling solution becomes unstable.  In particular, we find that this mode has an approximate wavelength, $l_{2}\simeq l_1/2$, where $l_1(=L/2)$ is the wavelength of the unstable mode at buckling of the vertical equilibrium.  Given that the critical force for the classical buckling of an Euler beam is inversely proportional to square of the buckling wavelength, it is reasonable that the dimensionless critical follower force for the instability of whirling may be estimated as $f^{**}\approx 4 f^*(=141.2)$. This value of $f^{**}$ is fairly close to the measured values of $f^{**}$ across all $\alpha$ (see Figure \ref{fig:secondbifurcation}B), in particular $f^{**} = 137.3$ for $\alpha=44$.  This suggests that secondary buckling is a potential physical mechanism for this bifurcation.

%% file: part7.tex
\section{From beating to writhing}\label{sec:beatingtowrithing}

\begin{figure}[htpb]
    \centering
    \includegraphics[width=0.9\textwidth]{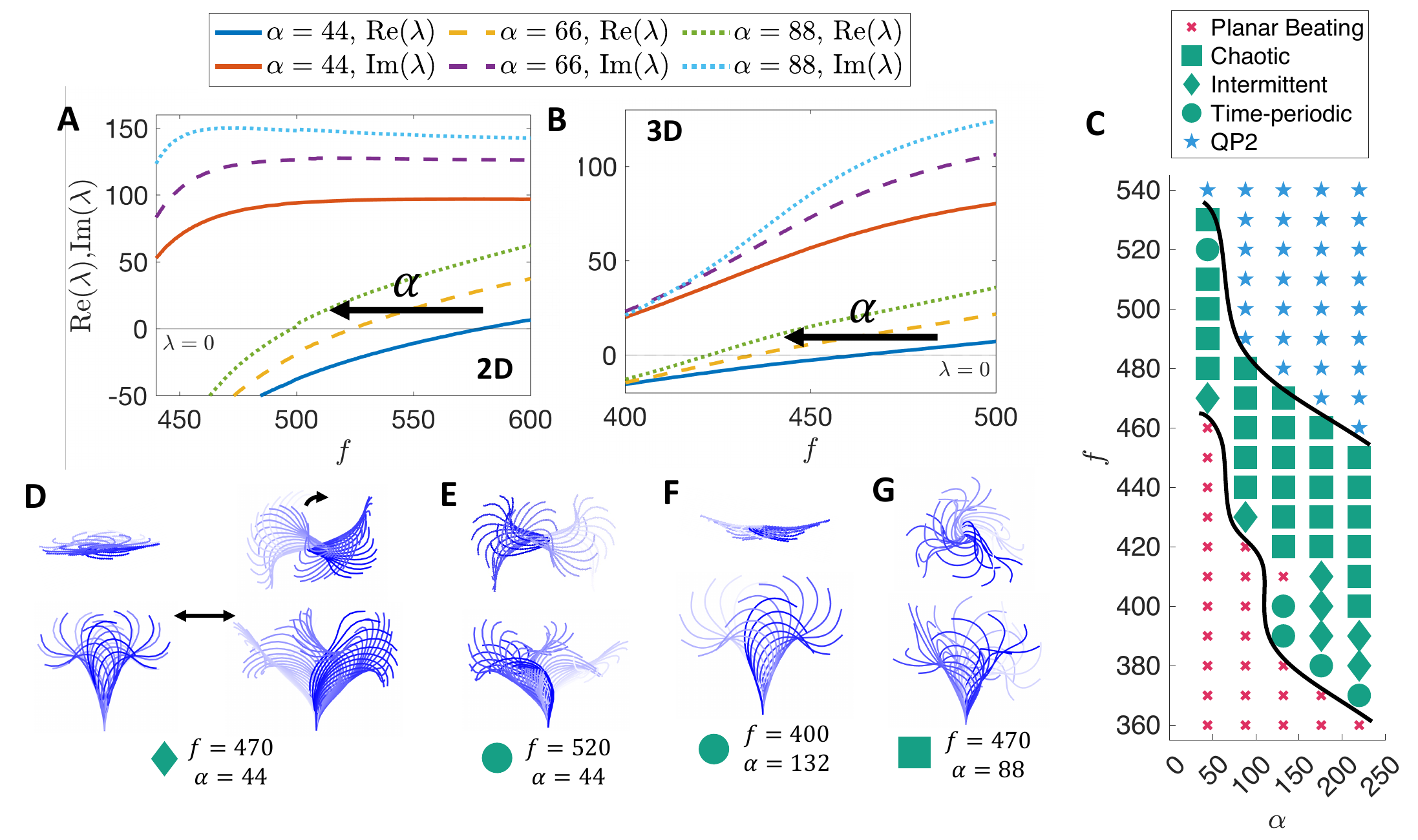}
   \caption{A: Eigenvalues from the 2D Floquet analysis of the planar beating state for $f\in [440,600]$ for different $\alpha$. Planar beating becomes unstable at $f = 578.9, 523.7$ and $498.1$ for $\alpha = 44, 66,$ and $88$, respectively. B: Eigenvalues from the 3D Floquet analysis of the planar beating state for $f\in [400,500]$ for different $\alpha$. The planar beating becomes unstable at $f = 462.8, 434.4$ and $422.4$ for $\alpha = 44, 66$ and $88$, respectively. C: The transition region between the planar beating (red crosses) and QP2 (blue stars) solutions.  Solutions in the region are divided into three types: chaotic (green squares), intermittent (green diamonds), and time-periodic (green circles). D-G: Filament motion in the transition region as viewed from the side and above for four of the observed behaviours. The passage of time is indicated by an increase in line boldness. D: Intermittence between beating with a growing out-of-plane component (left) and a combination of beating and whirling near the base and waving at the tip (right). E: Periodic solution with beating at the base and waving at the tip. F: Periodic solution where the filament beats while bent in the direction normal to the beat plane. G: Chaotic solution corresponding to erratic beating at the filament tip.}
    \label{fig:transition}
\end{figure}

By solving initial value problems, we have shown that beating is unstable at high follower force values, with the transition occurring at smaller $f$ as $\alpha$ increases.  This occurs both when the dynamics are restricted to 2D (Figure \ref{fig:picturesosfar}B), as well as in 3D (Figure \ref{fig:picturesosfar}C).  This picture is confirmed by the Floquet analysis of the beating solution, shown in Figure \ref{fig:transition}A, where we see that the critical follower force in 2D decreases from $f \approx 578.9$ for $\alpha = 44$ to $f \approx 498.1$ for $\alpha = 88$.  In 3D, we have $f \approx 462.8$ for $\alpha = 44$ decreasing to $f \approx 422.4$ for $\alpha = 88.$ (Figure \ref{fig:transition}B).  The Floquet analysis further reveals that at this bifurcation, in both 2D and 3D, the eigenvalues are imaginary.  Focusing on the fully 3D dynamics and revisiting the initial value problem just above the critical follower force, we observe that the solution enters a transition region that contains a whole host of behaviours including intermittency, periodic motion and chaotic beating as shown in Figure \ref{fig:transition}C. The intermittent dynamics are a subset of the chaotic solutions, exhibiting irregular bursts of nearly periodic motion interrupted by chaotic beating (see Figure \ref{fig:transition}D). At $(f,\alpha) = (410,132)$, we observe a transient period of intermittency that ultimately returns to the planar beating solution. 

These numerical experiments reveal that the observed dynamics are highly sensitive to $f$ and $\alpha$.  Immediately after the bifurcation we observe intermittency for lower $\alpha$.  For high $\alpha$, however, we find a periodic solution where the filament is beating while bent in the direction perpendicular to the beat plane (shown in Figure \ref{fig:transition}F). Additionally, we find that filaments with a smaller $\alpha$ can produce solutions, such as the periodic state shown in Figure \ref{fig:transition}E, that are not observed by filaments with a larger $\alpha$, and vice versa.  In general, at higher forces values, a greater variety of periodic solutions are observed.  For instance, as shown in Figure \ref{fig:transition}E, the filament base, broadly speaking, beats in one plane, while the tip oscillates in the transverse plane.  Near the final bifurcation (the upper black line in Figure \ref{fig:transition}C), we observe chaotic beating where the filament is rotating near the base while the tip waves aperiodically (Figure \ref{fig:transition}G). 

\begin{figure}
    \centering
    \includegraphics[width=\textwidth]{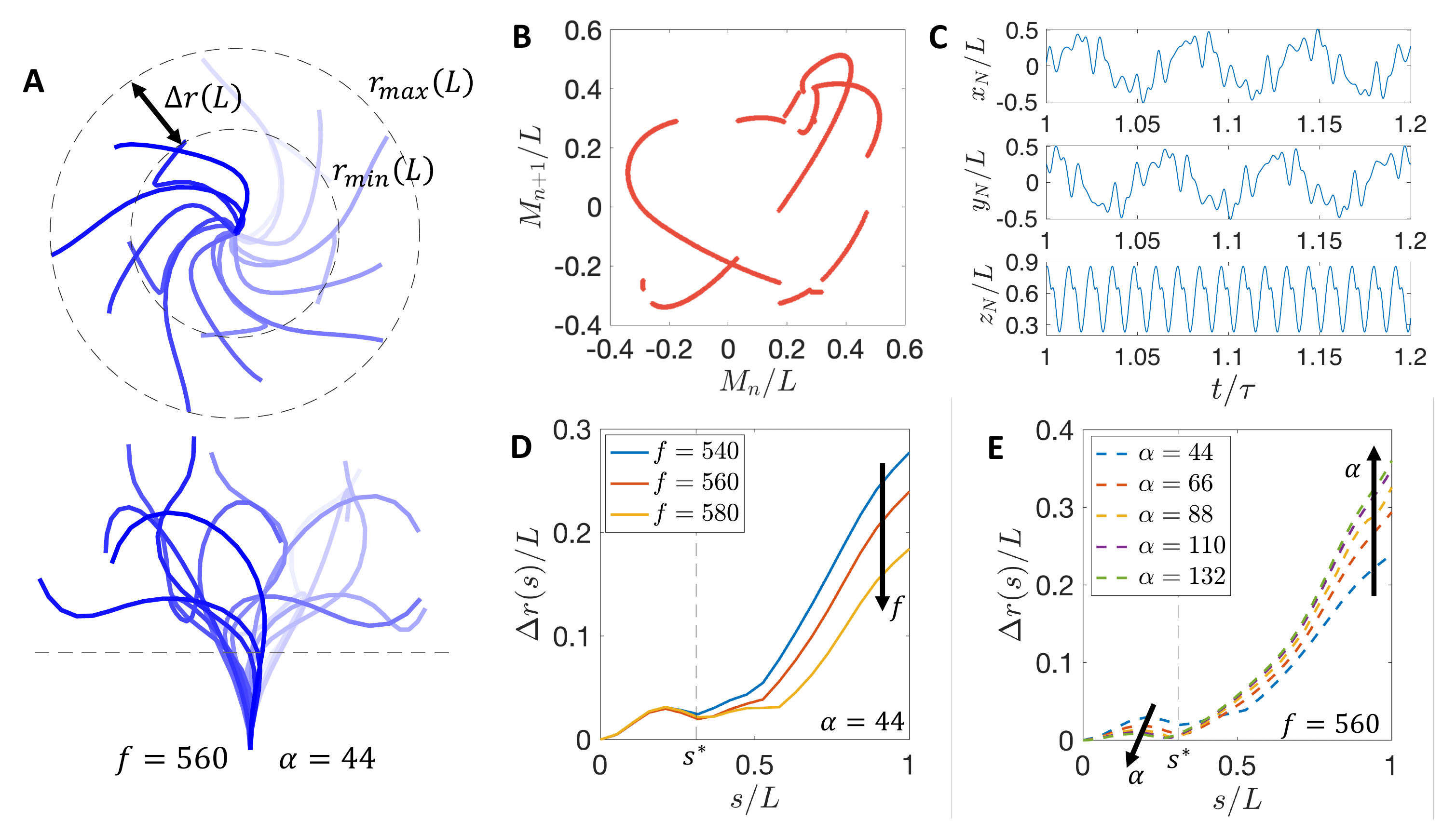}
    \caption{A: Filament motion in the QP2 regime.  The passage of time is indicated by an increase in line boldness. Top: A view from above showing the dynamics projected in the $(x,y)-$plane. The dashed lines indicate the minimum and maximum horizontal distance at $s=L,$ and $\Delta r(s) = r_{max}(s)-r_{min}(s).$  Below: A view from the side showing the dynamics projected in the $(x,z)-$plane. The dashed line corresponds to $s/L=s^*\approx 0.3$ from panels D and E. B: The Poincar\'{e} section corresponding to the QP2 solution for $f=560$ and $\alpha = 44$. C: The time-series of the filament tip displacement for $f=560$ and $\alpha = 44$ showing periodic oscillations in the $z$ component and irregular oscillations in the $x,y-$ components. D: The variation in radius, $\Delta r(s)/L$, against the arc-length, $s/L,$ for $\alpha=44$ and various $f$. E: The variation in radius, $\Delta r(s)/L$, as a function of $s/L$ for $f=560$ and various $\alpha$.}
    \label{fig:WhirlingWriggling}
\end{figure}

We find that this transition region exists for a finite range of the forcing values. Regardless of filament aspect ratio all filaments undergo the same quasiperiodic behaviour at the highest force values.  We refer to this solution as QP2.  The transition to QP2 occurs at lower force values as $\alpha$ increases.  For the QP2 solutions, the bottom half of the filament appears to be whirling while the upper half beats with a different frequency (see Figure \ref{fig:WhirlingWriggling}A). We find that QP2 arises for the variety of initial conditions that we have explored, including those generated by perturbing the trivial equilibrium, the asymptotically stable state in 2D, and the unstable whirling state.  This implies that that QP2 is presumably a stable attractor robust to perturbations and initial configurations.  To analyse the nature of this solution, we compute its Poincar\'{e} section, which reveals several disconnected curves (see Figure \ref{fig:WhirlingWriggling}B), confirming the quasiperiodicity of the solution.

To further investigate the effect of the forcing and aspect ratio on QP2, we compute the horizontal distance from the vertical equilibrium,
\begin{equation}
    r(s,t) = ||(\bm{I} - \hat{\bm{z}}\hat{\bm{z}}^T)\bm{Y}(s,t)||.
\end{equation}
After allowing the solution to reach its asymptotically stable state, we determine the minimum and maximum horizontal distance over time, $r_{min}(s) = \text{min}_{t\geq 0} \;r(s,t)$ and $ r_{max}(s) = \text{max}_{t \geq 0} \; r(s,t)$, respectively, and plot their difference $\Delta r(s) = r_{max}(s) - r_{min}(s)$. The results are shown for fixed aspect ratio and forcing in Figures \ref{fig:WhirlingWriggling}D and \ref{fig:WhirlingWriggling}E, respectively.  For small values of the arc-length, i.e. for  $s/L$ below $s^*$ (the local minimum of $\Delta r(s)$ for $(f,\alpha)=(560,44)$), the difference in the horizontal distance remains small, corresponding to whirling.  Above $s^*$, the difference rapidly increases as the beating amplitude increases with $s$. Increasing the follower force decreases the difference in the radius near the filament tip (Figure \ref{fig:WhirlingWriggling}D), while increasing the aspect ratio has the opposite effect (Figure \ref{fig:WhirlingWriggling}E).  Additionally, the local maximum in $\Delta r$ appearing at $s/L < s^*$ decreases in value as the aspect ratio increases. This implies that as the filament becomes more slender, the horizontal distance approaches a constant value.  Hence, the motion near the filament base more closely resembles whirling as $\alpha$ increases, where each segment trajectory lies on a circle of constant radius. Although QP2 is quasiperiodic, meaning the dynamics are a combination of multiple behaviours with periods that are not integer divisors, we can analyse how varying $(f,\alpha)$ affect the dynamics temporally by comparing dominant frequencies using the fast Fourier transform. See Appendix \ref{appen:extrafigs} for more details.

%% file: discussion.tex
\section{Discussion}
In this paper, we performed a thorough computational bifurcation analysis of the follower force model for an active filament clamped to a no-slip wall. Along with verifying the findings of \cite{DeCanio2017,Ling2018Instability-drivenMicrofilaments}, we have extended these studies to pinpoint and classify bifurcations, and identify new solutions.  By performing a linear stability analysis of the trivial state, we have established the presence of a double Hopf bifurcation at buckling, reconciling previous results from 2D and 3D studies of the model and establishing the connection between the eigenmodes in 2D with those in 3D. Turning our attention to the transition between whirling and planar beating occurring at higher force values, we show that there is not only one, but rather two, bifurcations that occur, with a new, stable quasiperiodic solution (QP1) that exists for a small range of $f$ between the two bifurcations.  Finally, we examine the solutions at high force values.  Here, we show there are two distinct regions: a transition region where a range of periodic and chaotic behaviours are observed, and a second quasiperiodic regime (QP2) where the filament exhibits rotation at the base and waving at the tip.  In performing these analyses, we have also established the role of $\alpha,$ the filament aspect ratio, providing a full picture of the parameter space governing the solutions that may arise.  In particular, we have shown how increasing $\alpha$ expands the QP1 region, decreases the force where beating becomes unstable, and produces different solutions observed in the transition regime.

It is interesting to note that while it is reasonable to assume MTs are actuated by more than one motor protein at any given time, the follower force model predicts that only a single motor protein is required to induce buckling, producing the filament motion explored in this work. The length of MTs are variable, but can be between $1\mu m - 100\mu m$ \cite{Bray2000CellMotility,Pampaloni2006ThermalLength,Ling2018Instability-drivenMicrofilaments}.  Taking the typical MT length to be $L \approx 20 \mu m$, the MT bending modulus as $K_B \approx 10^{-23} Nm^2$ \cite{Gittes1993FlexuralShape}, and the motor forcing as $6pN$ (molecular motors exert forces ranging between $1-10pN$ \cite{Schliwa2003MolecularMotors,Shingyoji1998DyneinGenerators}), the corresponding non-dimensional follower force is $f \approx 240$.  This value lies squarely in the planar beating regime, firmly suggesting that a single dynein provides sufficient force to cause the onset of self-sustained MT oscillations.  Due to the large distribution of parameter values reported experimentally, it is likely that even higher values of the non-dimensional forcing could also be attained.

While the follower force model considered in this paper provides the simplest representation of a motor protein driven filament, alternative models that provide more features of the biophysical system have been proposed. 
These include models with follower forces distributed along the entire filament length, or further, the direct effect of the motors on the surrounding fluid, either through an equal and opposite force on the surrounding fluid \cite{DeCanio2019,DeCanio2017,Stein2021SwirlingCytoskeleton}, or through a slip flow on the filament surface \cite{Laskar2017FilamentNumber}. Previous studies \cite{DeCanio2019,Ling2018Instability-drivenMicrofilaments} indicate that the distributed follower force case has qualitatively similar dynamics to the single follower force at the tip, also exhibiting whirling and beating behaviours as we vary the forcing. Moreover, the initial buckling from the vertical equilibrium, to whirling in the 3D case and beating in 2D, appears to also be a double Hopf bifurcation. The same argument from Section \ref{sec:doublehopf} could apply in this case to explain the connection between 2D and 3D buckling, suggesting that this initial buckling bifurcation may be generic to follower-force driven filament systems. Other variants of the model discussed in this work consider how changing the boundary conditions enforced on the filament can affect the resulting dynamics, for instance clamping both ends ($s=0$ and $s=L$) a distance $l < L$ apart \cite{Fatehiboroujeni2018NonlinearDrag,Fatehiboroujeni2021Three-dimensionalFlipping}. In these cases buckling still occurs, but the resulting dynamics are different to those observed for the clamped and free conditions we consider throughout this paper.

Fully classifying the emerging dynamics for variants of this model, and comparing the possible differences in the bifurcation diagrams could allow for an effective qualitative comparison of these models with actual MT dynamics and hence, better ascertain the correct models for predicting more complicated MT-motor protein related phenomena, such as collective dynamics.  As MT-motor protein complexes often arise in localised groups, hydrodynamic interactions between neighbouring filaments couples their motion, leading to important biological phenomena, such as cytoplasmic streaming.  The model including both follower forces distributed along the filament length and the opposite forces exerted on the fluid has been used to replicate this streaming, yielding a nontrivial steady state where all filaments are bent in the same direction if the filament density is sufficiently large \cite{Stein2021SwirlingCytoskeleton}. 

The follower force model can also be extended to describe more organised collections of MTs, such as the axoneme within cilia.  While the follower force does not replicate the shear-driven actuation in cilia, the dynamics exhibited by the basic follower force model studied in this paper is qualitatively similar to that observed for cilia.  Nodal cilia that play a vital role in symmetry breaking during embryogenesis undergo tilted whirling, pumping fluid unidirectionally \cite{Nonaka1998RandomizationProtein,SmithFluidCilia,Smith2011MathematicalCilia}.  Motile cilia elsewhere, such as respiratory cilia which use this motion to transport mucus in the lungs \cite{Chilvers2000AnalysisMethods}, typically perform planar beating.  Consequently, some studies employ a follower force model as a surrogate model to examine related ciliary motions, for instance coordination on spherical surfaces \cite{Westwood2021CoordinatedSurfaces}, or the effect of fluid rheology on the pumping ability of cilia \cite{Wang2023Generalized-NewtonianFilament,Link2023EffectFilament}.  Recent studies have incorporated the follower force model into an axoneme-like configuration of filaments, where equal and opposite follower forces are exerted on neighbouring filaments \cite{Bayly2016SteadyFlagella,Woodhams2022GenerationModels} to demonstrate that oscillations can arise by instabilities similar to those studied in this paper.  This provides an additional mechanism for cilia actuation alongside active dynein force regulation based on the geometric clutch hypothesis \cite{Lindemann1994AActivation,Lindemann1994AFlagella,Lindemann2002GeometricBeat}, local curvature of the axoneme \cite{Sartori2016DynamicFlagella,Gallagher2023AxonemalModulation,Dillon2000AnBeating,Dillon2003MathematicalMotility,Dillon2007FluidBeating}, or via other feedback mechanisms \cite{Chakrabarti2019SpontaneousMicrofilaments,Chakrabarti2021ACilia,Chakrabarti2022ACilia,Han2018SpontaneousCilia,Woodhams2022GenerationModels}.  A comparison between the bifurcations that arise in these different models could provide insight into the actuation process of cilia. In particular, applying the computational tools employed in this paper to models with different regulatory feedback loops, or indeed with no such feedback mechanism, will help to uncover whether active feedback between axoneme geometry and motor activity is required to reproduce experimental results.  This important direction, as well as the more fundamental studies of follower force model variants will be the focus of our future work.

%
%


%% file: appendix.tex
\appendix

\section{Hydrodynamics - the RPY tensor}\label{appen:RPY}
Due to the linearity of the Stokes equations, the mobility tensor in Eq \ref{eq:mobility_matrix_OG} can be decomposed into two parts, corresponding to the mobility tensor for the collection of particles in an unbounded domain, superposed with the corrections due to the presence of the wall \cite{Swan2007SimulationBoundary}. We can write this explicitly, by first using Eq \ref{eq:mobility_matrix_OG} to express the mobility for particle $n$ as:
\begin{equation}
    \left( \begin{array}{cc}
          \bm{V}_n \\ \bm{\Omega}_n
    \end{array}\right) = \sum_{m=1}^{N} \left(\begin{array}{cc}
         \bm{M}_{nm}^{VF} & \bm{M}_{nm}^{VT}  \\ \bm{M}_{nm}^{\Omega F} & \bm{M}_{nm}^{\Omega T} 
    \end{array}\right) \left( \begin{array}{cc}
         \bm{F}_m \\ \bm{T}_m 
    \end{array} \right).
\end{equation}
where $\bm{M}_{nm}^{VF}, \bm{M}_{nm}^{VT}, \bm{M}_{nm}^{\Omega F}$ and $\bm{M}_{nm}^{\Omega T}$ are $3 \times 3$ sub-tensors. As the mobility matrix is symmetric, we have $\bm{M}_{nm}^{VT} = \left(\bm{M}_{nm}^{\Omega F}\right)^T$.

The sub-tensors associated to the particle's self-mobility (i.e. for $n=m$) are given by \cite{Swan2007SimulationBoundary}:
\begin{align}
    \left(\bm{M}^{VF}_{nn}\right)_{ij} &= \frac{1}{6 \pi \eta a} \left( \delta_{ij} - \frac{1}{16}\left( \frac{9}{h_n} - \frac{2}{h_n^3} + \frac{1}{h_n^5}  \right)(\delta_{ij} - \delta_{i3}\delta_{j3}) - \frac{1}{8}\left( \frac{9}{h_n} - \frac{4}{h_n^3} + \frac{1}{h_n^5}  \right) \delta_{i3}\delta_{j3} \right) \\
    \left(\bm{M}_{nn}^{\Omega F}\right)_{ij} &= \frac{1}{64 \pi \eta a h_n^4} \epsilon_{3ji}\\
    \left(\bm{M}_{nn}^{\Omega T}\right)_{ij} &= \frac{1}{8\pi \eta a^3} \delta_{ij} - \frac{1}{64\pi\eta a h_n^3}\left(\frac{5}{2}(\delta_{ij} - \delta_{i3}\delta_{j3}) + \delta_{i3} \delta_{j3}  \right)
\end{align}
where $h_n = \bm{Y}_n \cdot \hat{\bm{e}}_z / a$ is the height of the particle above the wall, normalised by the particle's radius. The first terms for $\left(\bm{M}^{VF}_{nn}\right)_{ij} $ and $\left(\bm{M}_{nn}^{\Omega T}\right)_{ij}$ correspond to the contributions considering the particles in unbounded flow, and the subsequent terms arise from considerations of the wall. Terms from the final sub-block, $\left( \bm{M}_{nn}^{VT}\right)_{ij}$, can be obtained using symmetry.

To express the tensors corresponding to particle-particle interactions ($n\neq m$), we first define terms related to the difference in displacements of particles $n$ and $m$,
$\bm{Y}_{nm} =\bm{Y}_n - \bm{Y}_m$, $r_{nm} = ||\bm{Y}_{nm}|| $ and $\hat{\bm{Y}}_{nm}=(\hat{Y}_{nm}^1,\hat{Y}_{nm}^2,\hat{Y}_{nm}^3) = \bm{Y}_{nm}/r_{nm}$, as well as terms related to the difference in displacements of particle $n$ and the reflected image of particle $m,$ $\bm{R} = (\bm{Y}_n-\bm{Y}_m+2h_m\delta_3)/a,$ $R = ||\bm{R}||$ and $\bm{e}=\bm{R}/R$. Also defining $\hat{h} = h_m/(a\bm{R}\cdot \hat{\bm{e}}_z)$, these interaction terms are given by \cite{Swan2007SimulationBoundary}:
\begin{align}
    \begin{split}
    \left(\bm{M}^{VF}_{nm}\right)_{ij} = {} & \frac{1}{8 \pi \eta r_{nm}}\left(\left(1+\frac{2a^2}{3r^2_{nm}}\right) \delta_{ij} + \left( 1 - \frac{2a^2}{r_{nm}^2}\hat{Y}_{nm}^i\hat{Y}_{nm}^j\right) \right)\\   
    & +\frac{1}{12 \pi \eta a}\left[ -\frac{1}{2}\left(\frac{3\left(1+2 \hat{h}(1-\hat{h}) e_3^2\right)}{R}+\frac{2\left(1-3 e_3^2\right)}{R^3} - \frac{2\left(1-5 e_3^2\right)}{R^5}\right) \delta_{i j}\right.\\
    &\left.-\frac{1}{2}\left( \frac{3\left(1-6 \hat{h}(1-\hat{h}) e_3^2\right)}{R}-\frac{6\left(1-5 e_3^2\right)}{R^3}+ \frac{10\left(1-7 e_3^2\right)}{R^5}\right) e_i e_j \right.\\
    &\left.+ e_3\left( \frac{3 \hat{h}\left(1-6(1-\hat{h}) e_3^2\right)}{R}- \frac{6\left(1-5 e_3^2\right)}{R^3}+ \frac{10\left(2-7 e_3^2\right)}{R^5}\right) e_i \delta_{j 3} \right.\\
    &\left. + e_3\left(\frac{3 \hat{h}}{R}-\frac{10}{R^5}\right) \delta_{i 3} e_j -2\left(\frac{3 \hat{h}^2 e_3^2}{R}+\frac{3 e_3^2}{R^3}+\frac{\left(2-15 e_3^2\right)}{R^5}\right) \delta_{i 3} \delta_{j 3} \right],
    \end{split} \\
    \begin{split}
    \left(\bm{M}_{nm}^{\Omega F}\right)_{ij} = {}& \frac{1}{8 \pi \eta r_{nm}^2}\epsilon_{ijk} \hat{Y}_k - \frac{1}{4\pi\eta a} \left(\frac{1}{2R^2} \epsilon_{i j k} e_k+\left(\frac{6 \hat{h} e_3^2}{R^2}+\frac{\left(1-10 e_3^2\right)}{R^4}\right) \epsilon_{3 k i} e_k \delta_{j 3} \right. \\
    & \left. - e_3\left(\frac{3 \hat{h}}{R^2}-\frac{5}{R^4}\right) \epsilon_{3 k i} e_k e_j- e_3\left(\frac{\hat{h}}{R^2}-\frac{1}{R^4}\right) \epsilon_{3 i j}\right),
    \end{split}\\
    \begin{split}
    \left(\bm{M}_{nm}^{\Omega T}\right)_{ij} = {}& \frac{1}{16 \pi \eta r_{nm}^3} \left( 3\hat{Y}_{nm}^i \hat{Y}_{nm}^j - \delta_{ij}\right) + \frac{3}{8\pi \eta a}\left( \frac{\left(1-6 e_3^2\right)}{6R^3} \delta_{i j} -\frac{1}{2R^3} e_i e_j \right. \\
    & \left. +\frac{e_3}{R^3} \delta_{i 3} e_j+\frac{1}{R^3} \epsilon_{3 k i} \epsilon_{3 l j} e_k e_l\right).
    \end{split}
\end{align}
The first term in each expression originates from considerations of the hydrodynamics in an unbounded domain, while the following terms correspond to the presence of the no-slip wall, as derived in \cite{Swan2007SimulationBoundary}. Again, the terms from the final sub-block, $\left( \bm{M}_{nm}^{VT}\right)_{ij}$, can be obtained using symmetry.

\section{The Relaxation Time}\label{appen:relaxation_timescale}

\begin{figure}
    \centering
    \includegraphics[width=\textwidth]{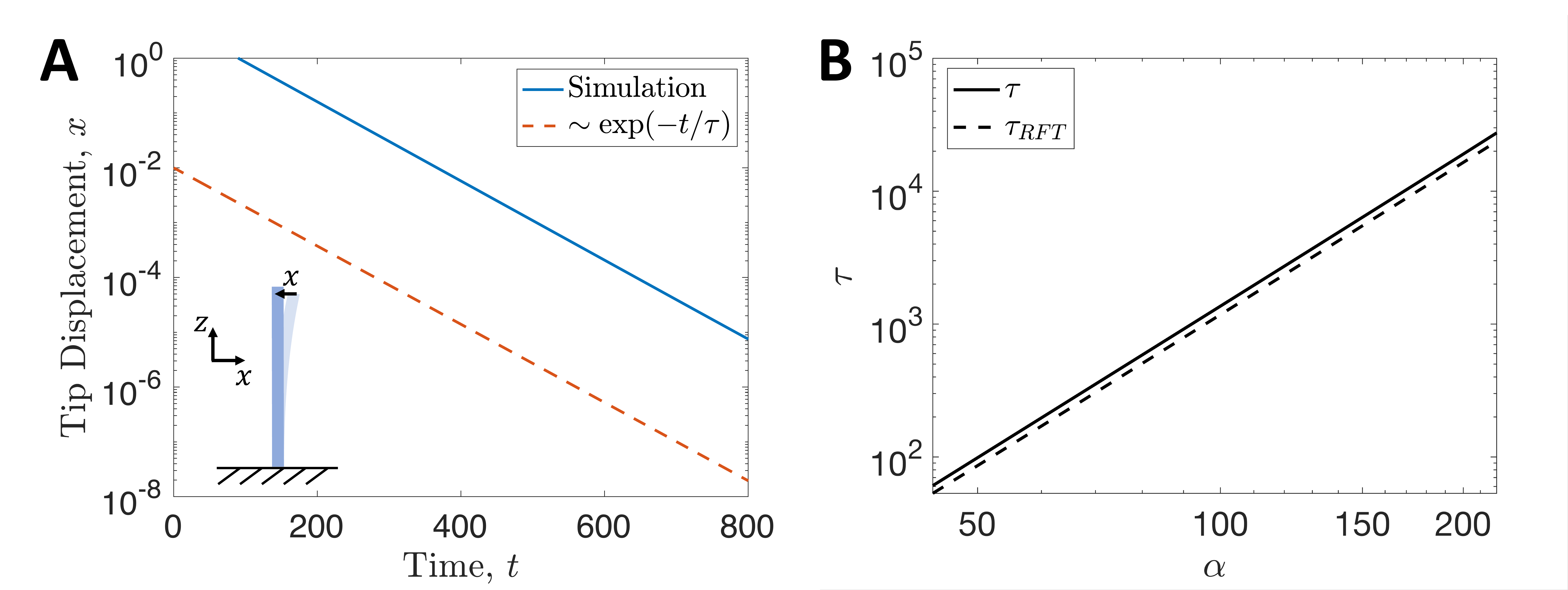}
    \caption{A: The tip displacement $x,$ (solid line) as a function of time for $\alpha = 44$.  The dashed line $\sim \exp(-t/\tau)$, where $\tau$ is the relaxation timescale determined through fitting the displacement data.  The schematic depicts how the computation of the relaxation timescale is performed. B: The relaxation timescale computed numerically for our model as a function of aspect ratio, compared to the timescale arising from resistive force theory.}
    \label{fig:time_discretisation}
\end{figure}

Due to the dependence of the relaxation time on the hydrodynamic model and also the boundary conditions on the fluid domain, we calculate the relaxation time, $\tau,$ numerically. To do this, we measure the tip-displacement of a force-free filament as it relaxes back to the vertical equilibrium from a deformed state. As the initial deformation is small, we are in the linear regime and so expect the tip displacement to decay like $\sim \exp(-t/\tau)$.  After using this to fit the decay data, we can extract $\tau$ (see Figure \ref{fig:time_discretisation}).  As this timescale depends on filament aspect ratio, $\alpha$, this process must be repeated for filaments with different aspect ratio.

From \cite{Wiggins1998FlexiveNumber}, the relaxation time for a filament is 
\begin{equation}
    \tau_{RFT} = \frac{\zeta_{\perp}L^4}{(1.875)^4K_B}
\end{equation}
where $\zeta_{\perp}$ is the perpendicular drag per unit length given by resistive force theory \cite{Keller1976Slender-bodyFlow},
\begin{equation}
    \zeta_{\perp} = \frac{4 \pi \eta}{\log(2L/a) - 1/2}.
\end{equation}
Comparing with $\tau$ given by our model, we find a similar dependence on the filament aspect ratio with the approximate relation $\tau \approx 1.1 \tau_{RFT}$ (see Figure \ref{fig:time_discretisation}).

\section{Eigenvalues of the Transition Operator}\label{appen:arnoldi}
In order to establish the stability of steady and time-periodic states, we compute the eigenvalues of $\bm{\Phi}(T;\bm{U}_0)$, where $T$ is an arbitrary, but short, time interval for a steady solution, or the time period for a periodic solution.  The matrix $\bm{\Phi}(t;\bm{U}_0)$ is defined by $\delta\bm{U}(t)=\bm{\Phi}(t;\bm{U}_0)\delta \bm{U}(0)$. To compute its eigenvalues, we combine our numerical solver, which advances the filament configuration in time, with the Arnoldi method.  The effect of multiplying a vector $\delta \bm{U}(0)$, with $||\delta \bm{U}(0)|| = \epsilon \ll 1$, by the state transition operator for the time interval $T$, is approximately given by
\begin{align}
    \bm{\Phi}(T;\bm{U}_0) \delta\bm{U}(0) &  = \delta \bm{U}(T) \\ 
    & = \bm{\varphi}(\bm{U}_0(0) + \delta \bm{U}(0),T) - \bm{U}_0(T) \\
    & = \bm{\varphi}(\bm{U}_0(0) + \delta \bm{U}(0),T) - \bm{U}_0(0),
\end{align}
where the first equality is the definition of $\bm{\Phi}(T;\bm{U}_0)$, and the second equality follows from the definition of $\bm{\varphi}$, Eq (\ref{eq:varphi_definition}), for $\bm{U} = \bm{U}_0 + \delta \bm{U}$. The final equality follows as $\bm{U}_0$ is either steady, or time-periodic with period $T$. Defining $\hat{\bm{x}} = \delta\bm{U}/\epsilon$, we can apply the Arnoldi method using the approximation of the matrix-vector multiplication between the transition operator and a unit vector, $\hat{\bm{x}}$,
\begin{equation}
    \bm{\Phi}(T;\bm{U}_0) \hat{\bm{x}} \approx \frac{1}{\epsilon} \left( \varphi(\bm{U}_0 + \epsilon \hat{\bm{x}},T) - \bm{U}_0 \right).
\end{equation}
This is equivalent to solving the initial value problem with initial condition $\bm{U}_0 + \epsilon \hat{\bm{x}}$, extracting the configuration after a time $T$, subtracting the steady or time-periodic configuration, $\bm{U}_0$, and dividing by the magnitude of the perturbation, $\epsilon$.

At iteration $k$, the Arnoldi method decomposes the transition operator into 
\begin{equation}
\bm{\Phi}(T;\bm{U}_0)\approx \bm{Q}_{k+1} \bm{H}_k \bm{Q}_k^*.
\end{equation}
Here, $\bm{Q}_{k+1} \in \mathbb{C}^{3(N-1)\times  (k+1)}$ is an orthogonal matrix, the column space of which spans a Krylov subspace of $\bm{\Phi}(T;\bm{U}_0)$, and $\bm{H}_k \in \mathbb{C}^{(k+1)\times k}$ is an upper Hessenberg matrix. For large $k$, the eigenvalues of $\tilde{\bm{H}}_k$, the first $k\times k$ sub-block of $\bm{H}_k$, approximate those of $\bm{\Phi}(T;\bm{U}_0)$. As the extremal eigenvalues of $\tilde{\bm{H}}_k$ will be the first to converge, the most unstable mode can be determined directly from $\tilde{\bm{H}}_k$ after several iterations when the matrix is typically small in size and standard eigenvalue solvers can be used.  The eigenvalues of $\tilde{\bm{H}}_k$ (say $\mu$) approximate those of $\bm{\Phi}(T;\bm{U}_0)$. Then, we can determine the eigenvalues, say $\lambda$, of $\bm{A}$ or $\bm{B}$ through
\begin{equation}
 \lambda = \frac{1}{T} \log (\mu),
\end{equation}
following from (\ref{eq:phi_steady}) and (\ref{eq:phi_timeperiodic}).  The corresponding eigenmode can then be computed by applying $\bm{Q}_{k}$ to the eigenvector of $\tilde{\bm{H}}_k$.

\section{Eigenmodes of the Unsteady Whirling State}\label{appen:eigenmodes}
\begin{figure}
    \centering
    \includegraphics[width=\textwidth]{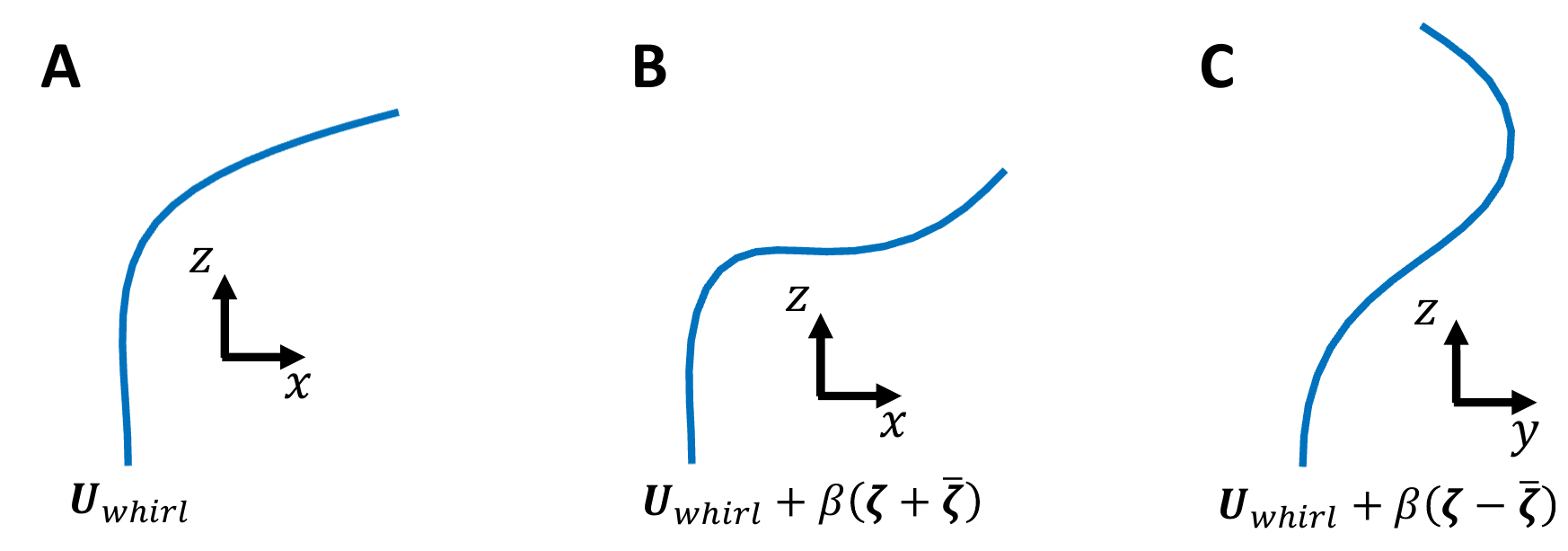}
    \caption{The filament shapes for A: the whirling solution, $\bm{U}_{\text{whirl}}$ for $f=137.4$ and $\alpha =44$ and B/C: the sum of the whirling solution and a linear combination of the unstable eigenvectors,  $\bm{U}_{\text{whirl}} + \beta ( \bm{\zeta} \pm \bar{\bm{\zeta}})$, for $\beta = 3/2$.}
    \label{fig:eigenmodes}
\end{figure}

Floquet analysis revealed that the whirling state becomes unstable at $f \approx 137.2$ for $\alpha = 44$ (see Figure \ref{fig:secondbifurcation}B). Our linear stability analysis yields the eigenvalues and eigenvectors for each state, which we can then use to provide insight into the nature of the instability. At $f=137.4$ (i.e. soon after the bifurcation) and $\alpha = 44$, we see the whirling solution has two unstable eigenvalues, $\lambda,\bar{\lambda},$ and two corresponding eigenvectors, $\bm{\zeta},\bar{\bm{\zeta}}.$

In Figure \ref{fig:eigenmodes}, we plot $\bm{U}_{\text{whirl}} + \beta_1 \bm{\zeta} + \beta_2 \bar{\bm{\zeta}}$, the sum of the whirling solution, $\bm{U}_{\text{whirl}}$ (left panel) and a linear combinations of the two associated unstable eigenmodes, for the cases $\beta_1 = \beta_2 = 3/2 $ (centre panel), and $\beta_1 = -\beta_2 = 3/2$ (right panel). From these figures, we can see that the wavelength stemming from the eigenmode is approximately half that of the wavelength of the unstable mode at buckling.

\section{Effect of Twist}\label{appen:twist}
In this work, we set the bending and twisting moduli to be equal. The appropriate choice of $\gamma = K_T/K_B$ in our simulations is non-trivial as these parameters are difficult to measure experimentally and their values appearing in the literature can vary significantly. Even the parameters which can be used to estimate these moduli, such as the Young's modulus and shear modulus, have experimental estimates spanning several orders of magnitude \cite{Feng2018Data-drivenMechanics,Kis2002NanomechanicsMicrotubules}.

For the purpose of this work, we can generate a rough estimate of these parameters by making the assumption that the MT is isotropic. Then, the shear modulus can be expressed in terms of the Young's modulus, $E$, and Poisson's ratio, $\nu$, by $\mu = E/2(1+\nu)$. Experimentally, the Young's modulus of MTs is suggested to lie in the range $3.1 \text{ MPa} - 1.4 \text{ GPa}$ \cite{Feng2018Data-drivenMechanics}, and, following \cite{Sirenko1996ElasticFiuid}, we can estimate Poisson's ratio as $\nu = 0.3$. This gives an estimated range of the shear modulus: $\mu \in (1.2,540) \text{ MPa}$. By approximating the MT as a cylindrical shell with internal/external diameters $2r_i = 15 \text{nm}$ and $2r_e = 25 \text{nm}$ respectively \cite{Safinya2019MinireviewBiomolecules}, the bending and twisting moduli are then given by \cite{Landau1986TheoryEdn}
\begin{equation}
    K_B = \frac{ \pi E}{4} \left(r_e^4 - r_i^4  \right), \;\;\;
    K_T = \frac{ \pi \mu}{2} \left(r_e^4 - r_i^4  \right),
\end{equation}
respectively. Using our parameters, this gives broad estimates in the range $K_B \in (5.2 \times 10^{-26},2.3\times 10^{-23}) \text{Nm}^2$ and $K_T \in (4.0 \times 10^{-26},1.8 \times 10^{-23}) \text{Nm}^2$. Experimental evidence suggests the bending modulus of MTs could vary between around $2 \times 10^{-24} \text{Nm}^2$ \cite{Felgner1996FlexuralTweezers} and $2 \times 10^{-23} \text{Nm}^2$ \cite{Gittes1993FlexuralShape}. This implies that $\gamma$ can range between approximately $1.7 \times 10^{-3}$ and $350$ for MTs; over five orders of magnitude.  In \cite{Cheminiak2010TorsionalModel}, calculations that take into account the anisotropy of MTs suggest this ratio to be around $\gamma \approx O(0.01)$.

\begin{figure}
    \centering
    \includegraphics[width=0.9\textwidth]{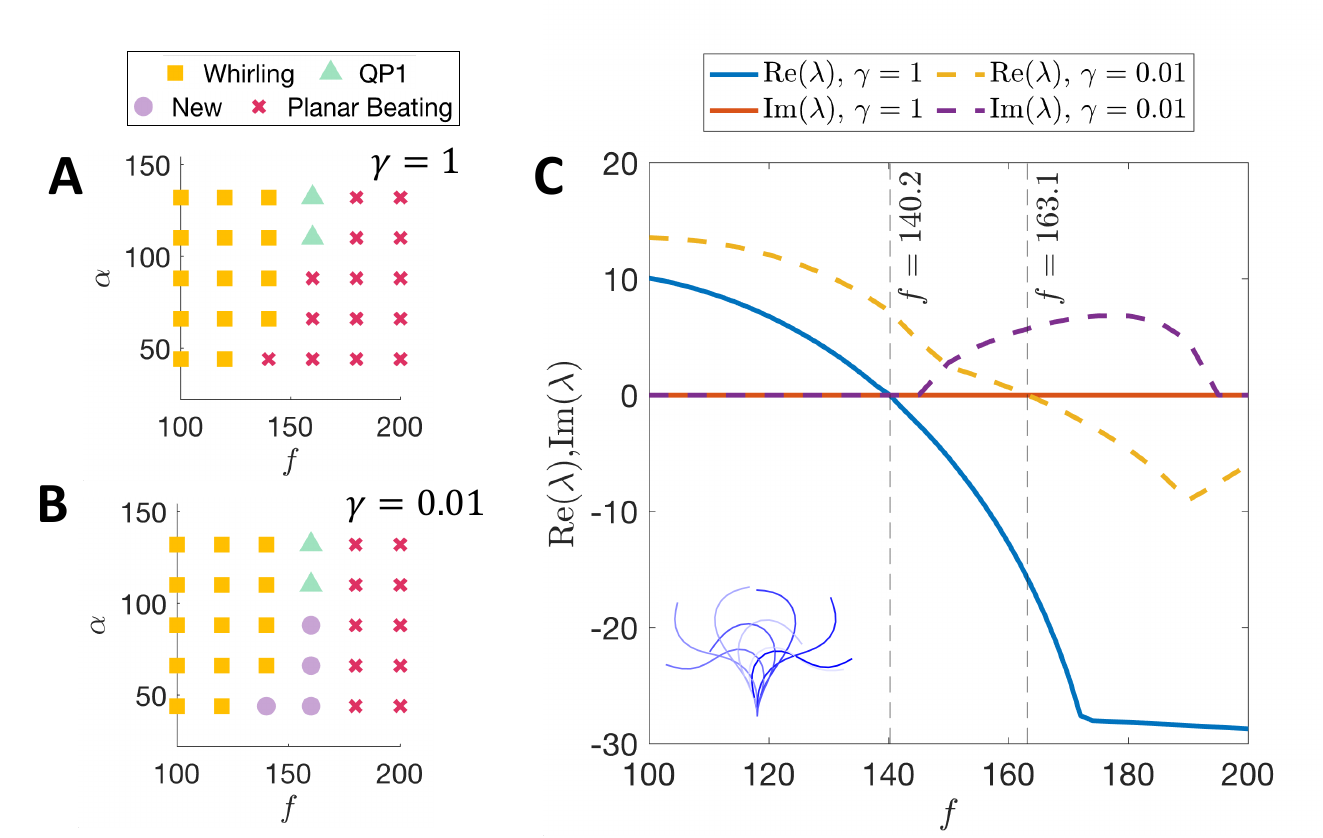}
    \caption{The phase diagram for $f\in (100,200),$ $ \alpha \in (44,132)$ for A: $\gamma=1$ and B: $\gamma=0.01$. C: Floquet analysis on the beating solution for a filament with aspect ratio $\alpha = 44$ for $\gamma \in \{0.01,1\}.$ We note that the bifurcation occurs later for $\gamma=0.01,$ and the eigenvalue is complex at the bifurcation.}
    \label{fig:twist}
\end{figure}

To investigate the effects of the twisting modulus on our model, we compare results for $\gamma \in \{0.01, 1, 100\}$. We run initial value problems for $f \in \{0,300\}$ and observe that the state space does not change when increasing from $\gamma = 1$ to $\gamma=100$. However for $\gamma=0.01,$ we observe differences for filaments with smaller aspect ratios -- planar beating becomes stable at a higher value of the forcing, as shown in Figure \ref{fig:twist}. In its place, we see a range of distinct dynamics to those observed for $\gamma=1$. For instance, for $(f,\alpha) = (140,44)$ the filament appears as a mixture of whirling, QP1 and planar beating. The filament begins by whirling, and then transitions to planar beating through QP1 type-beating, before returning to whirling in the opposite direction. For $(f,\alpha)=(160,88)$ on the other hand, the filament undergoes periodic planar deformations.

By performing Floquet analysis on the planar beating behaviours for $\gamma=0.01$, we can confirm the beating becomes stable for higher forcing values, as indicated by the simulations. Although the dynamics are different, for $\alpha \in \{66,88\}$, the dominant eigenmode remains purely real at this bifurcation, as for $\gamma=1$. However, for $\alpha=44,$ the eigenvalue is complex as shown in Figure \ref{fig:twist}, indicating a change in frequency between the new dynamics and the planar beating at this transition.

These results are not entirely unexpected.  The torsional drag scales like $\sim O(\alpha^{-3})$, and so increasing the aspect ratio will decrease the amount of torsion generated. Therefore it is expected that changing the twist modulus should not change the results significantly, and any changes will occur for smaller aspect ratio, as observed and discussed above.

\section{QP2 Frequencies}\label{appen:extrafigs}
\begin{figure}
    \centering
    \includegraphics[width=0.5\textwidth]{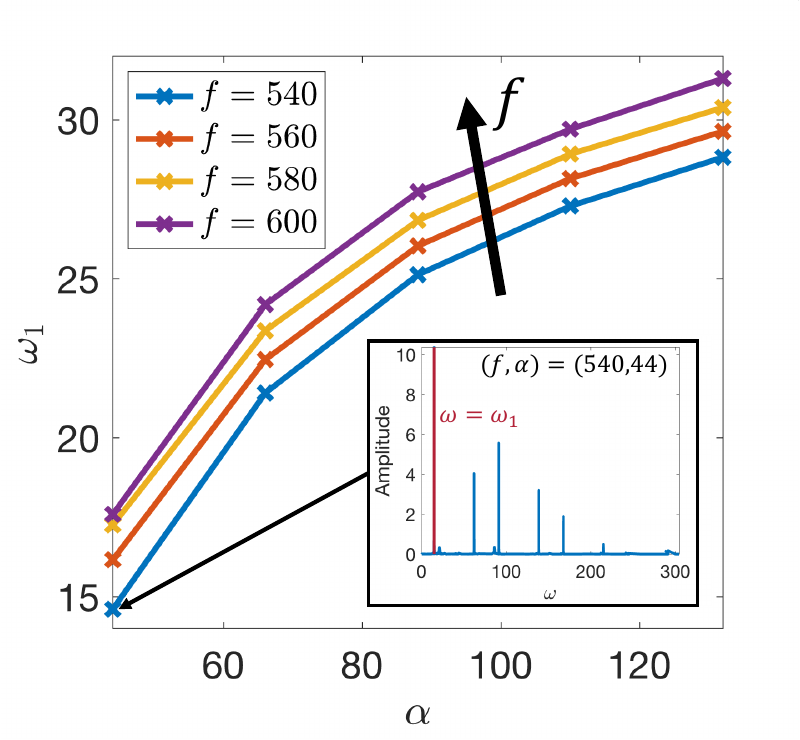}
    \caption{The frequency associated to the dominant mode, $\omega_1$, for various $f$ and $\alpha.$ We see that $\omega_1$ increases with both $f$ and $\alpha$. Inset: Single sided amplitude spectrum of QP2 for $(f,\alpha) = (540,44).$ We take $\omega_1$ to be the frequency with the largest amplitude. }
    \label{fig:extrafig_fft}
\end{figure}

The solution that emerges for the highest forcing values, QP2, is quasiperiodic, meaning it is a composition of several periods which are not integer divisors. In particular, this means we cannot extract a frequency in the usual sense. Instead, in order to compare these dynamics as we vary the forcing and aspect ratio, we use a fast Fourier transform (FFT) on the $x-$coordinate of the tip to decompose our dynamics into their corresponding frequencies (see Figure \ref{fig:extrafig_fft} inset). We can extract the frequency with the most dominant amplitude (i.e. the largest spike) and plot these for various $f$ and $\alpha$, as shown in Figure \ref{fig:extrafig_fft}. As with the planar beating and whirling solutions (as shown in Figure \ref{fig:secondbifurcation}A), we observe that this frequency increases with both forcing and aspect ratio. Although these results are obtained by decomposing the time-trace of the tip $x-$coordinate, we note that the dominant frequencies are found to be the same when analysing the displacement of any other segment, indicating that this frequency is likely associated to the dynamics of the lower portion of the filament. Recall that in QP2, the base of the filament is whirling whereas the upper portion appears to be beating, while being `dragged' around by the whirling base. It appears as though the mode corresponding to this whirling is the dominant contribution to the dynamics along the filament length. 

\section{Supplementary Videos}
We include the below supplemental videos which display the key dynamics discussed in the main text. 

\textbf{SM1} (\textit{sm1\_whirl.mp4}): A filament with $(f,\alpha) = (100,44)$, i.e. undergoing whirling oscillations.

\textbf{SM2} (\textit{sm2\_qp1.mp4}): A filament with $(f,\alpha) = (139,44)$, i.e. undergoing QP1 oscillations.

\textbf{SM3} (\textit{sm3\_beat.mp4}): A filament with $(f,\alpha) = (300,44)$, i.e. undergoing planar beating oscillations.

\textbf{SM4} (\textit{sm4\_qp2.mp4}): A filament with $(f,\alpha) = (600,44)$, i.e. undergoing QP2 oscillations.